\documentclass{article}

\usepackage{amssymb,latexsym,amsmath}

\usepackage[pdftex]{graphicx}

\usepackage{hyperref}

\begin{document}

\pagestyle{plain}

\newtheorem{theorem}{Theorem}[section]

\newtheorem{proposition}[theorem]{Proposition}

\newtheorem{lema}[theorem]{Lemma}

\newtheorem{corollary}[theorem]{Corollary}

\newtheorem{definition}[theorem]{Definition}

\newtheorem{remark}[theorem]{Remark}

\newtheorem{exempl}{Example}[section]

\newenvironment{exemplu}{\begin{exempl}  \em}{\hfill $\square$

\end{exempl}}

\newcommand{\ea}{\mbox{{\bf a}}}

\newcommand{\eu}{\mbox{{\bf u}}}

\newcommand{\ueu}{\underline{\eu}}

\newcommand{\ueo}{\overline{u}}

\newcommand{\oeu}{\overline{\eu}}

\newcommand{\ew}{\mbox{{\bf w}}}

\newcommand{\ef}{\mbox{{\bf f}}}

\newcommand{\eF}{\mbox{{\bf F}}}

\newcommand{\eC}{\mbox{{\bf C}}}

\newcommand{\en}{\mbox{{\bf n}}}

\newcommand{\eT}{\mbox{{\bf T}}}

\newcommand{\eL}{\mbox{{\bf L}}}

\newcommand{\eR}{\mbox{{\bf R}}}

\newcommand{\eV}{\mbox{{\bf V}}}

\newcommand{\eU}{\mbox{{\bf U}}}

\newcommand{\ev}{\mbox{{\bf v}}}

\newcommand{\eve}{\mbox{{\bf e}}}

\newcommand{\uev}{\underline{\ev}}

\newcommand{\eY}{\mbox{{\bf Y}}}

\newcommand{\eK}{\mbox{{\bf K}}}

\newcommand{\eP}{\mbox{{\bf P}}}

\newcommand{\eS}{\mbox{{\bf S}}}

\newcommand{\eJ}{\mbox{{\bf J}}}

\newcommand{\eB}{\mbox{{\bf B}}}

\newcommand{\eH}{\mbox{{\bf H}}}

\newcommand{\leb}{\mathcal{ L}^{n}}

\newcommand{\eI}{\mathcal{ I}}

\newcommand{\eE}{\mathcal{ E}}

\newcommand{\hen}{\mathcal{H}^{n-1}}

\newcommand{\eBV}{\mbox{{\bf BV}}}

\newcommand{\eA}{\mbox{{\bf A}}}

\newcommand{\eSBV}{\mbox{{\bf SBV}}}

\newcommand{\eBD}{\mbox{{\bf BD}}}

\newcommand{\eSBD}{\mbox{{\bf SBD}}}

\newcommand{\ecs}{\mbox{{\bf X}}}

\newcommand{\eg}{\mbox{{\bf g}}}

\newcommand{\paromega}{\partial \Omega}

\newcommand{\gau}{\Gamma_{u}}

\newcommand{\gaf}{\Gamma_{f}}

\newcommand{\sig}{{\bf \sigma}}

\newcommand{\gac}{\Gamma_{\mbox{{\bf c}}}}

\newcommand{\deu}{\dot{\eu}}

\newcommand{\dueu}{\underline{\deu}}

\newcommand{\dev}{\dot{\ev}}

\newcommand{\duev}{\underline{\dev}}

\newcommand{\weak}{\stackrel{w}{\approx}}

\newcommand{\mild}{\stackrel{m}{\approx}}

\newcommand{\lrightarrow}{\stackrel{L}{\rightarrow}}

\newcommand{\rrightarrow}{\stackrel{R}{\rightarrow}}

\newcommand{\strong}{\stackrel{s}{\approx}}

\newcommand{\weakdown}{\rightharpoondown}

\newcommand{\opg}{\stackrel{\mathfrak{g}}{\cdot}}

\newcommand{\opunu}{\stackrel{1}{\cdot}}
\newcommand{\opdoi}{\stackrel{2}{\cdot}}

\newcommand{\opn}{\stackrel{\mathfrak{n}}{\cdot}}
\newcommand{\opx}{\stackrel{x}{\cdot}}

\newcommand{\tr}{\ \mbox{tr}}

\newcommand{\Ad}{\ \mbox{Ad}}

\newcommand{\ad}{\ \mbox{ad}}

\renewcommand{\contentsname}{ }

\title{Graphic lambda calculus}

\author{Marius Buliga \\ 
\\
Institute of Mathematics, Romanian Academy \\
P.O. BOX 1-764, RO 014700\\
Bucure\c sti, Romania\\
{\footnotesize Marius.Buliga@imar.ro}}

\date{This version: 23.05.2013}

\maketitle

\begin{abstract}
We introduce and study graphic lambda calculus, a visual language which can be used for representing untyped lambda calculus, but it can also be used for computations in emergent algebras or for representing Reidemeister moves of locally planar tangle diagrams. 
\end{abstract}

\section{Introduction}

Graphic lambda calculus consists of a class of graphs endowed with moves between them. It might be considered a   visual language in the sense of Erwig \cite{erwig}. The name "graphic lambda calculus" comes from the fact that it can be used for representing terms and reductions from untyped lambda calculus. It's main move is called "graphic beta move" for it's relation to the beta reduction in lambda calculus. However, the graphic beta move can be applied outside the "sector" of untyped lambda calculus, and the graphic lambda calculus can be used for other purposes than the one of visual representing lambda calculus. 

For other visual, diagrammatic representation of lambda calculus see the VEX language \cite{vex}, or David Keenan's \cite{mock}. 

The motivation for introducing graphic lambda calculus comes from  the study of emergent algebras. In fact, my goal is to build eventually a logic system which can be used for the formalization of certain "computations" in emergent algebras, which can be applied then for a  discrete differential calculus which exists  for metric spaces with dilations, comprising riemannian manifolds and sub-riemannian spaces with very low regularity.

Emergent algebras are a generalization of quandles, namely an emergent algebra is a family of idempotent right quasigroups indexed by the elements of an abelian group, while quandles are self-distributive idempotent right quasigroups. Tangle diagrams decorated by quandles or racks  are a well known tool in knot theory \cite{fennrourke} \cite{joyce}. 

It is notable to mention the work of Kauffman \cite{kauf}, where the author uses knot diagrams for representing combinatory logic, thus untyped lambda calculus. Also  Meredith and Snyder\cite{mersny} associate to any knot diagram a process in pi-calculus, 

Is there any common ground between these three apparently separated field, namely differential calculus, logic and tangle diagrams? As a first attempt for understanding this, I proposed $\lambda$-Scale calculus  \cite{lambdascale}, which  is a formalism which contains both untyped lambda calculus and emergent algebras. Also,  in the paper \cite{buligachora} I proposed a formalism of decorated tangle diagrams for emergent algebras and I called "computing with space" the various manipulations of these diagrams with geometric content. Nevertheless, in that paper I was not able to give a precise sense of the use of the word "computing". I speculated, by using analogies from studies of the visual system, especially the "Brain a geometry engine" paradigm of Koenderink \cite{koen}, that, in order for the visual front end of the brain to reconstruct the visual space in the brain, there should be a kind of "geometrical computation" in the neural network of the brain akin to the manipulation of decorated tangle diagrams described in our paper. 

I hope to convince the reader that graphic lambda calculus gives a rigorous answer to this question, being a formalism which contains, in a sense, lambda calculus, emergent algebras and tangle diagrams formalisms.

\paragraph{Acknowledgement.} This work was supported by a grant of the Romanian National Authority for Scientific Research, CNCS – UEFISCDI, project number 
PN-II-ID-PCE-2011-3-0383.

\section{Graphs and moves}
\label{gra}

An oriented graph is a pair $(V,E)$, with $V$ the set of nodes and $E \subset V \times V$ the set of edges.  Let us denote by $\displaystyle \alpha: V  \rightarrow 2^{E}$ the map which associates to any node $N \in V$ the set of adjacent edges $\alpha(N)$. In this paper we work with locally planar graphs with  decorated nodes, i.e. we shall attach to a graph $(V,E)$ supplementary information: 
\begin{enumerate}
\item[-] a function $f: V \rightarrow A$ which associates to any node $N \in V$ an element 
of the "graphical alphabet" $A$ (see definition \ref{defalp}), 
\item[-] a cyclic order of $\alpha(N)$ for any $N \in V$, which is equivalent to giving a  local embedding of the node $N$ and edges adjacent to it into the plane. 
\end{enumerate}

We shall construct a set of locally  planar graphs with decorated nodes, starting from a graphical alphabet of  elementary graphs. On the set of graphs we shall define  local transformations, or moves.  Global moves or conditions will be then introduced. 

\begin{definition}
The graphical alphabet contains the elementary graphs, or gates, denoted by $\lambda$, $\Upsilon$, $\curlywedge$, $\top$, and for any element $\varepsilon$ of the commutative group $\Gamma$, a graph denoted by  $\bar{\varepsilon}$. Here are the elements of the graphical alphabet: 
\begin{enumerate}
\item[] $\lambda$ graph \hspace{1.cm} \includegraphics[width=20mm]{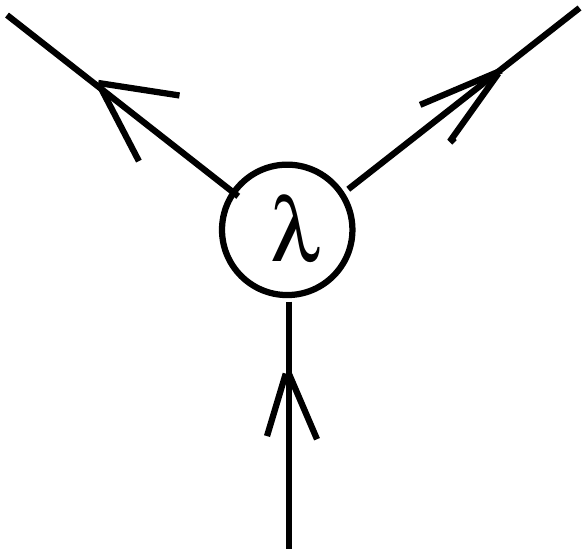}, \hspace{2.cm}
$\Upsilon$ graph \hspace{1.cm}\includegraphics[width=20mm]{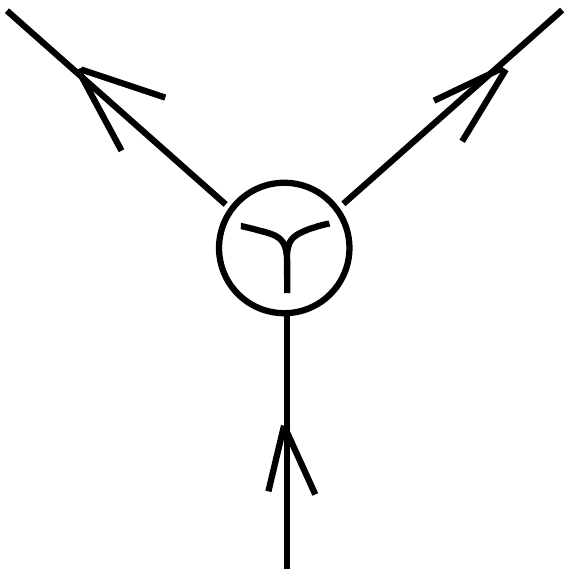}, 

\item[] $\curlywedge$ graph \hspace{1.cm}\includegraphics[width=15mm]{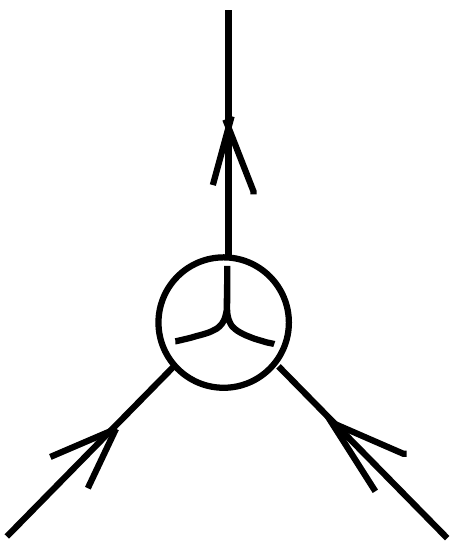}\hspace{1.cm}, 
\hspace{2.cm} $\bar{\varepsilon}$ graph \hspace{1.cm}\includegraphics[width=15mm]{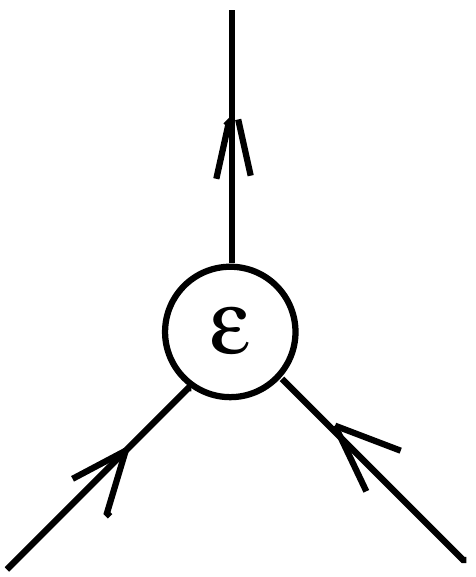}\hspace{.2cm}, 
\item[] $\top$ graph \hspace{1.cm}\includegraphics[width=8mm]{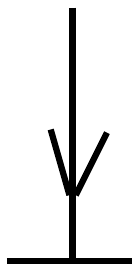}\hspace{1.cm}.
\end{enumerate}
With the exception of the $\top$, all other elementary graphs have three edges. The graph  $\top$ has only one edge. 
\label{defalp}
\end{definition}

There are two types of "fork" graphs, the $\lambda$ graph and the $\Upsilon$ graph, and two types of "join" graphs, the $\curlywedge$ graph and the $\bar{\varepsilon}$ graph. Further I briefly explain what are they supposed to represent and why they are needed in this graphic formalism. 

The $\lambda$ gate corresponds to the lambda abstraction operation from untyped lambda calculus. This gate has one input (the entry arrow) and two outputs (the exit arrows), therefore, at first view, it cannot be a graphical representation of an operation. In untyped lambda calculus the $\lambda$ abstraction operation has two inputs, namely a variable name  $x$ and a term $A$, and one output, the term $\lambda x . A$.   There is an algorithm, presented in section \ref{constru}, which transforms a lambda calculus term into a graph made by elementary gates, such that to any lambda abstraction which appears in the term corresponds a $\lambda$ gate. 

The $\Upsilon$ gate corresponds to a FAN-OUT gate. It is needed because the graphic lambda calculus described in this article does not have variable names. $\Upsilon$ gates appear in the process of elimination of variable names from lambda terms, in the algorithm previously mentioned. 

Another justification for the existence of two fork graphs is that they are subjected to different moves: the $\lambda$ gate appears in the graphic beta move, together with the $\curlywedge$ gate, while the $\Upsilon$ gate appears in the FAN-OUT moves. Thus, the $\lambda$ and $\Upsilon$ gates, even if they have the same topology, they are subjected to different moves, which in fact characterize their "lambda abstraction"-ness and the "fan-out"-ness of the respective gates.  The alternative, which consists into using only one, generic, fork gate, leads to the identification, in a sense, of lambda abstraction with  fan-out, which would be confusing. 

The $\curlywedge$ gate corresponds to the application operation from lambda calculus. The algorithm from  section \ref{constru} associates a $\curlywedge$ gate  to any application operation used in a lambda calculus term. 

The $\bar{\varepsilon}$ gate corresponds to an idempotent right quasigroup  operation, which appears in  emergent algebras, as an abstractization of the geometrical operation of taking a dilation (of coefficient $\varepsilon$), based at a point and applied to another point. 

As previously, the existence of two join gates, with the same topology, is justified by the fact that they appear in different moves.

\paragraph{1. The set GRAPH.} We construct the set of graphs $GRAPH$ over the graphical alphabet by grafting edges of a finite number of copies of the elements  of the graphical alphabet. 

\begin{definition} 
$GRAPH$ is the set of graphs obtained by grafting edges of a finite number of copies of the elements  of the graphical alphabet. During the grafting procedure, we start from a set of gates and we add, one by one, a finite number of gates, such that, at any step,  any edge of any elementary graph is grafted on any other free  edge (i.e. not already grafted to other edge) of the graph, with the condition  that they have  the same orientation.

For any node of the graph, the local embedding into the plane is given by the element of the graphical alphabet which decorates it. 

The set of free edges of a graph $G \in GRAPH$ is named the set of leaves $L(G)$. Technically, one may imagine that we complete the graph $G \in GRAPH$ by adding to the free extremity of any free edge a decorated node, called "leaf",  with decoration "IN" or "OUT", depending on the orientation of the respective free edge. The set of leaves $L(G)$ thus decomposes into a disjoint union $L(G) = IN(G) \cup OUT(G)$ of in or out leaves. 

Moreover, we admit into $ GRAPH$ arrows without nodes,          \includegraphics[width=15mm]{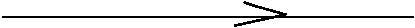} , called   wires  or lines,  and  loops  (without nodes from the elementary graphs, nor leaves)

 \vspace{.5cm}   \centerline{\includegraphics[width=20mm]{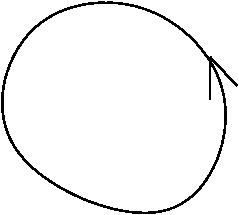}}  \vspace{.5cm}

Graphs in $GRAPH$ can be disconnected.  Any graph which is a finite reunion of lines, loops and assemblies of the elementary graphs  is in $ GRAPH$.

\end{definition}

\paragraph{2. Local moves.} These are transformations of graphs in $GRAPH$ which are local, 
in the sense that any of the  moves apply to a limited part of a graph, keeping the rest of the graph unchanged. 

We may define a local move as a rule of transformation of a graph into another of the following form.

First, a subgraph of a graph $G$ in $GRAPH$ is any collection of nodes and/or edges of $G$. It is not supposed that the mentioned subgraph must  be in $GRAPH$. Also, a collection of some edges of $G$, without any node, count as a subgraph of $G$. Thus, a subgraph of $G$ might be imagined as a subset of the reunion of nodes and edges of $G$. 

For any   natural  number $N$ and any graph $G$ in $GRAPH$, let  
$\displaystyle \mathcal{P}(G,N)$ be the collection  of subgraphs $P$ of the graph $G$ which have the sum of the number of  edges and nodes less than or equal to $N$.

\begin{definition}
A local move has the following form: there is a number $N$ and a condition $C$ which is formulated in terms of graphs which have the sum of the number of  edges and nodes less than or equal to $N$,  such that for any graph $G$ in $GRAPH$ and for any $P \in \mathcal{P}(G,N)$, if $C$ is true for $P$ then transform $P$ into $P'$, where $P'$ is also a graph which have the sum of the number of  edges and nodes less than or equal to $N$. 
\end{definition}

Graphically we may group the elements of the subgraph,  subjected to the application of the local rule,  into a region  encircled with a dashed closed, simple curve.  The edges which cross the curve (thus connecting the subgraph $P$ with the rest of the graph) will be numbered clockwise. The transformation will affect only the part of the graph which is inside the dashed curve (inside meaning the bounded connected part of the plane which is bounded by the dashed curve) and, after the transformation is performed, the edges of the transformed graph will connect to the graph outside the dashed curve by respecting the numbering of the edges which cross the dashed line. 

However, the grouping of the elements of the subgraph has no intrinsic meaning in graphic lambda calculus. It is just a visual help and it is not a part of the formalism.  As a visual help, I shall use sometimes colors in the figures. The colors, as well, don't have any intrinsic meaning in the graphic lambda calculus.

\paragraph{2.1. Graphic $\beta$ move.} This is the most important move, inspired by the $\beta$-reduction from lambda calculus, see theorem \ref{lambdathm}, part (d). 

\vspace{.5cm}   \centerline{ \includegraphics[width=100mm]{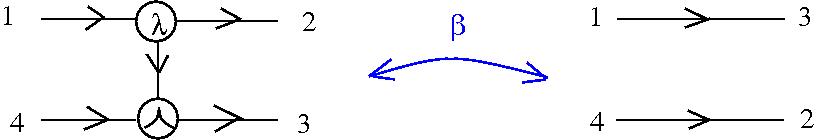}}  \vspace{.5cm}

The labels "1, 2, 3, 4" are used only as guides for gluing correctly the new pattern, after removing the old one. As with the encircling dashed curve, they have no intrinsic meaning in graphic lambda calculus. 

This "sewing braids" move will be used also in contexts outside of lambda calculus! It is the most powerful move in this graphic calculus. A primitive form of this move appears as the re-wiring move (W1) (section 3.3, p. 20  and the last paragraph and figure from section 3.4, p. 21 in \cite{buligachora}). 


An alternative notation for this move is the following:

\vspace{.5cm}   \centerline{ \includegraphics[width=100mm]{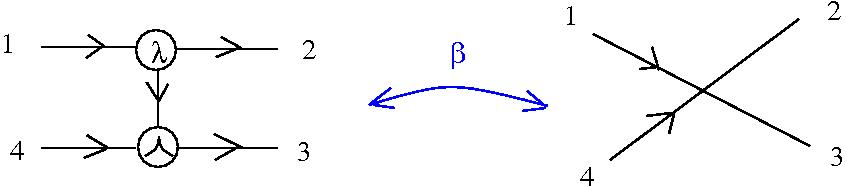}}  \vspace{.5cm}

A move which looks very much alike   the graphic beta move is the   UNZIP operation   from  the formalism of  knotted trivalent graphs, see for example the paper      \cite{thurs} section 3. 
In order to see this, let's draw again the graphic beta move, this time without labeling the arrows:

\vspace{.5cm}   \centerline{\includegraphics[width=120mm]{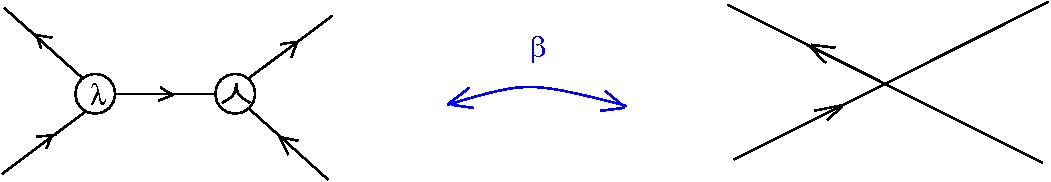}}  \vspace{.5cm}

The unzip operation acts only from left to right in the following figure. Remarkably, it acts on trivalent graphs (but not oriented).

\vspace{.5cm}   \centerline{\includegraphics[width=120mm]{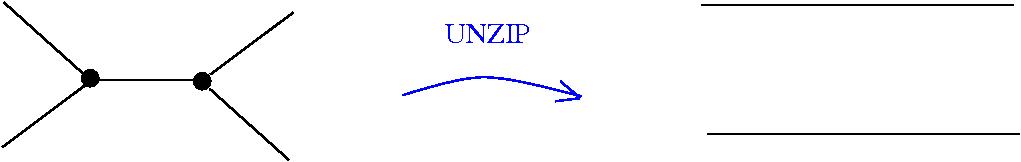}}  \vspace{.5cm}

Let us go back to the graphic beta move and remark that it does not depend on  the particular embedding in the plane. For example,  the intersection of the "1,3" arrow with the "4,2" arrow is an artifact of the embedding, there is no node there. Intersections of arrows have no meaning, remember that we work with graphs which are locally planar, not globally planar.

The graphic beta move goes into both directions. In order to apply the move, we may pick a pair of arrows and label them with "1,2,3,4", such that, according to the orientation of the arrows,  "1" points to "3" and "4" points to "2", without any node or label between "1" and "3" and between "4" and "2" respectively. Then, by a graphic beta move, we may replace the portions of the two arrows which are between "1" and "3", respectively between "4" and "2", by the pattern from the LHS of the figure. 

The graphic beta move may be applied even to a single arrow, or to a loop. In the next figure we see three  applications of the graphic beta move. They illustrate the need   for considering loops and wires as members of $ GRAPH$.

\vspace{.5cm}   \centerline{ \includegraphics[width=120mm]{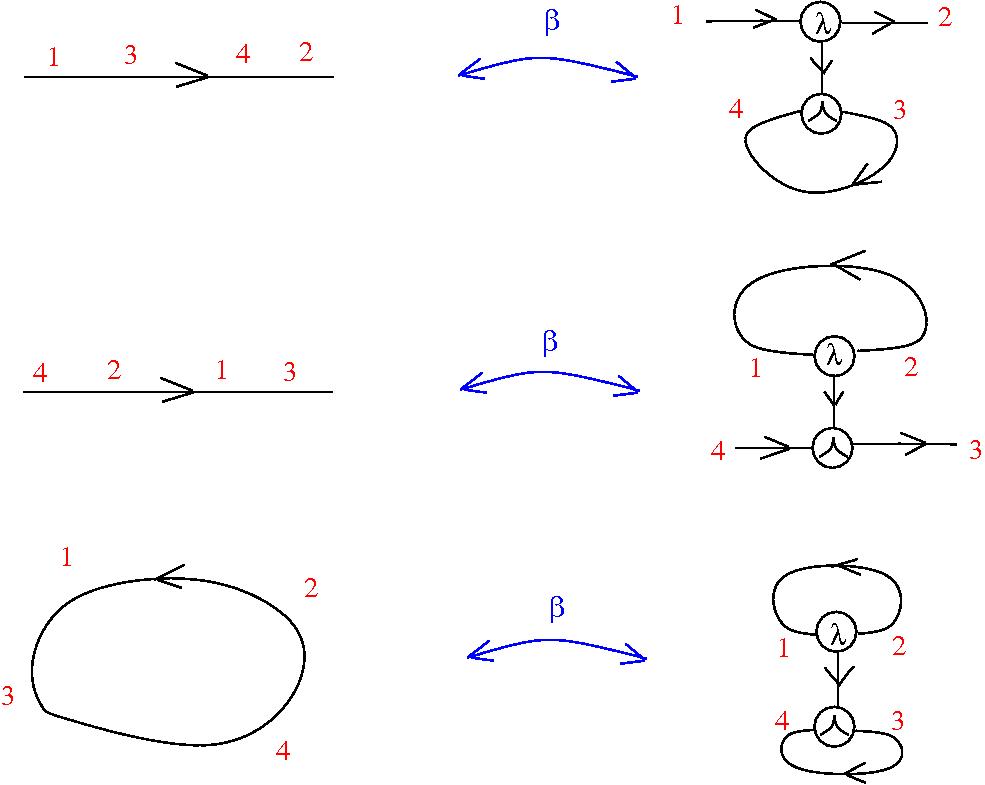}}  \vspace{.5cm}

Also, we can apply in different ways a graphic beta move, to the same graph and in the same place, simply by using different labels "1", ... "4" (here $A$, $B$, $C$, $D$ are graphs in $GRAPH$): 

\vspace{.5cm}   \centerline{ \includegraphics[width=135mm]{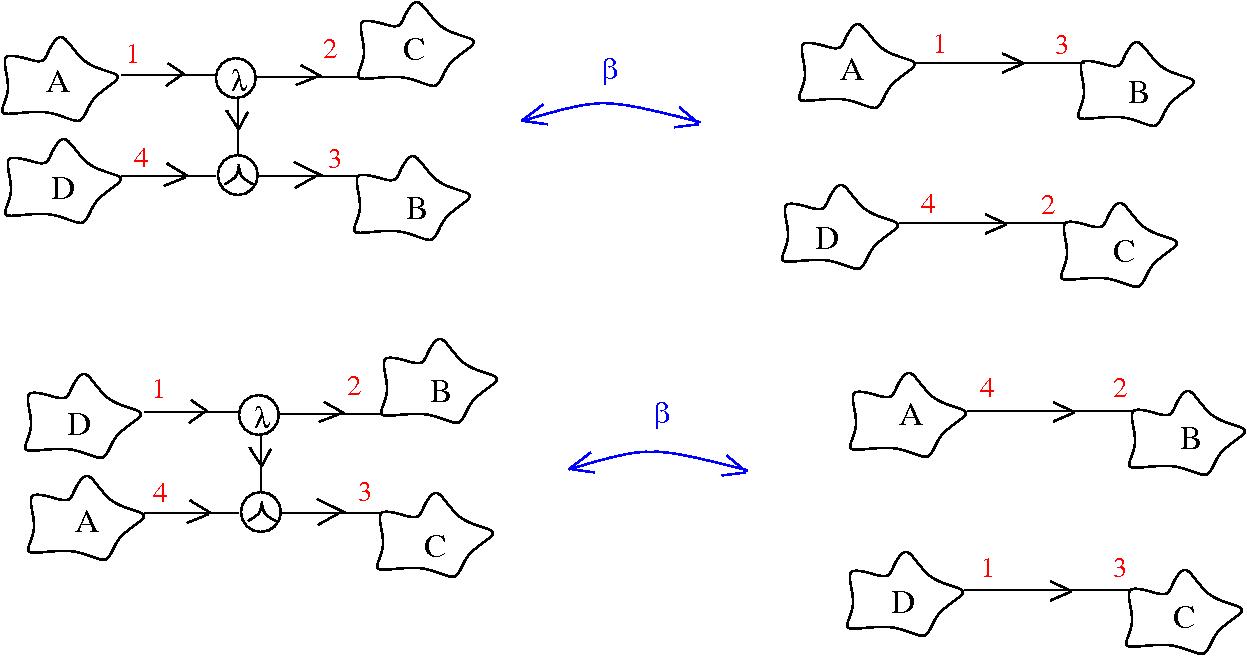}}  \vspace{.5cm}

A particular case of the previous figure is yet another justification for having loops as elements in $ GRAPH$.

\vspace{.5cm}   \centerline{\includegraphics[width=135mm]{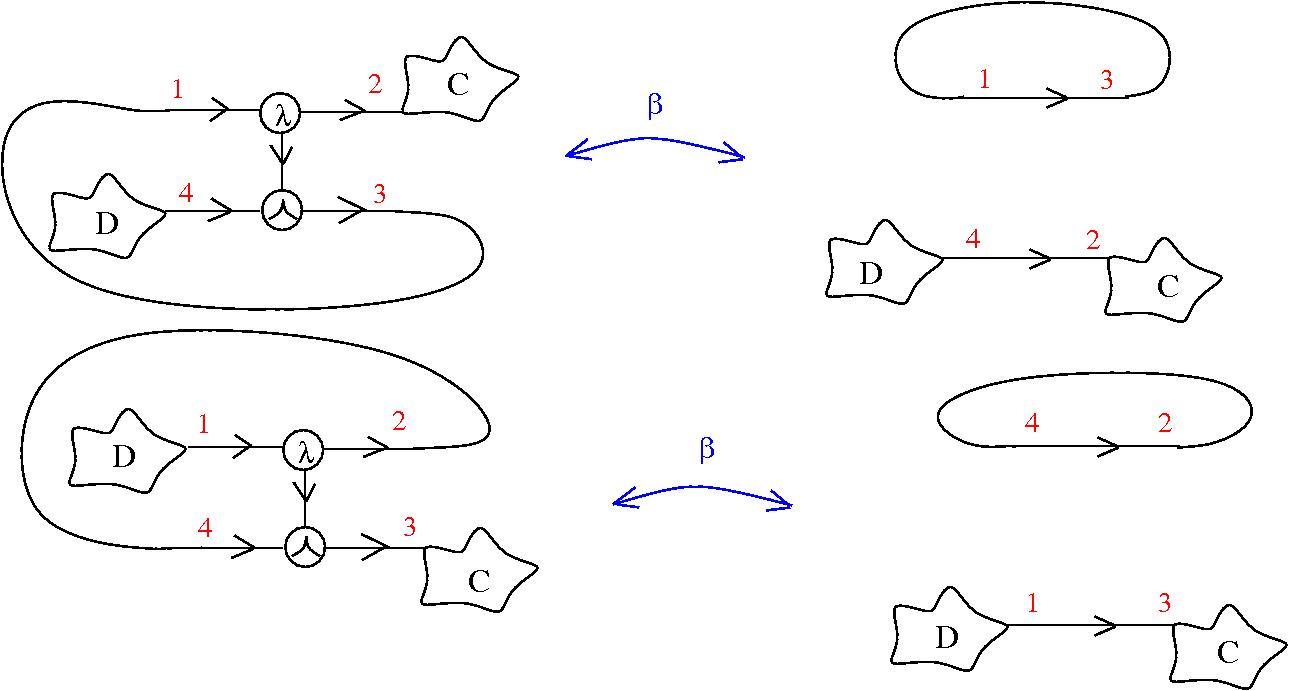}}  \vspace{.5cm}

These two applications of the graphic beta move may be represented alternatively like this:

\vspace{.5cm}   \centerline{\includegraphics[width=135mm]{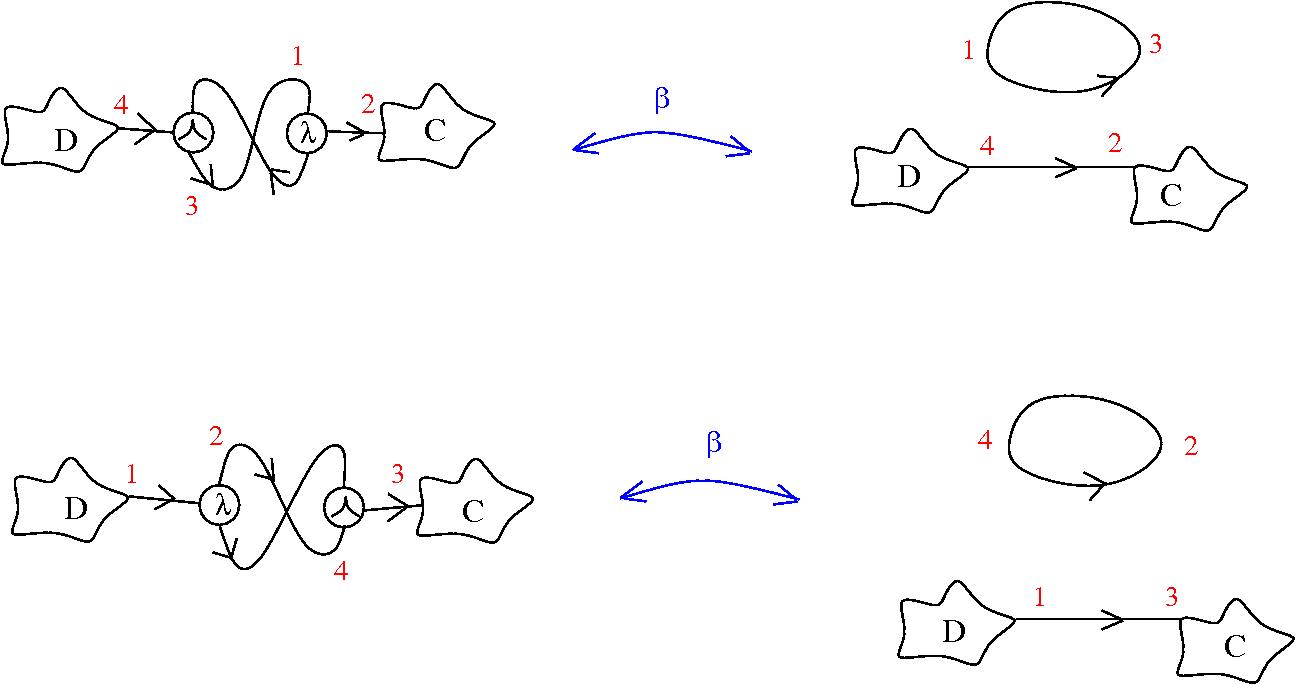}}  \vspace{.5cm}

\paragraph{2.2. (CO-ASSOC) move.} This is the "co-associativity" move involving the $\Upsilon$ graphs. We think about the  $\Upsilon$ graph as corresponding to a FAN-OUT gate. 

\vspace{.5cm}   \centerline{\includegraphics[width=80mm]{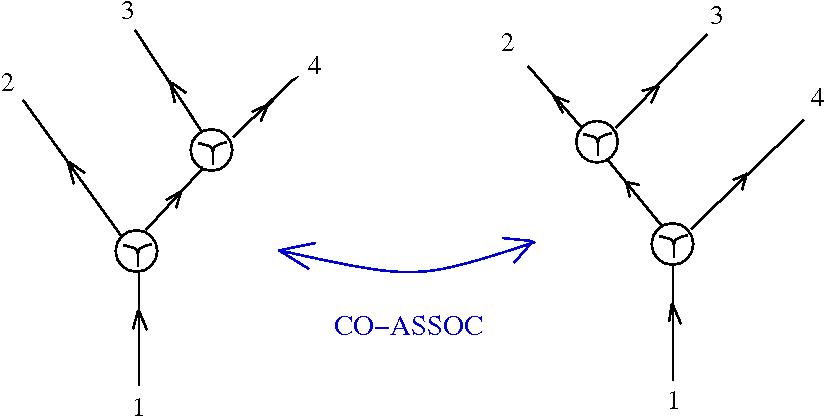}}  \vspace{.5cm}  

By using CO-ASSOC moves, we can move between any two binary trees formed only with $ \Upsilon$ gates, with the same number of output leaves. 

\paragraph{2.3. (CO-COMM) move.} This is the "co-commutativity" move involving the $\Upsilon$ gate. It will be not used until the section \ref{secbraid} concerning knot diagrams. 

\vspace{.5cm}   \centerline{\includegraphics[width=80mm]{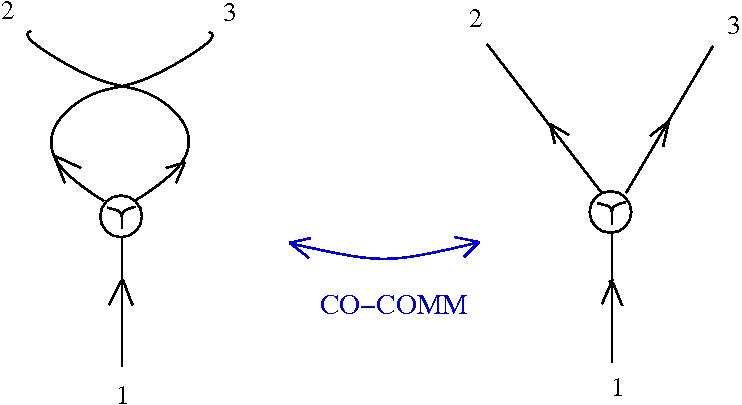}}  \vspace{.5cm}  

\paragraph{2.3.a (R1a) move.} This move is imported from emergent algebras. Explanations are given in section \ref{semer}.  It involves an $\Upsilon$ graph and a $\bar{\varepsilon}$ graph, with $\varepsilon \in \Gamma$. 

\vspace{.5cm}   \centerline{\includegraphics[width=100mm]{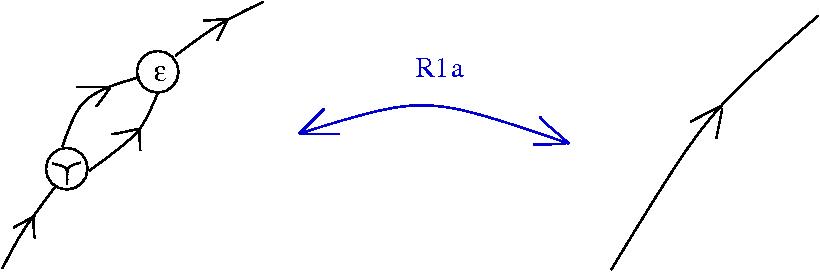}}  \vspace{.5cm}

\paragraph{2.3.b (R1b) move.}   The move R1b  (also related to emergent algebras) is this:   
   
\vspace{.5cm}   \centerline{\includegraphics[width=100mm]{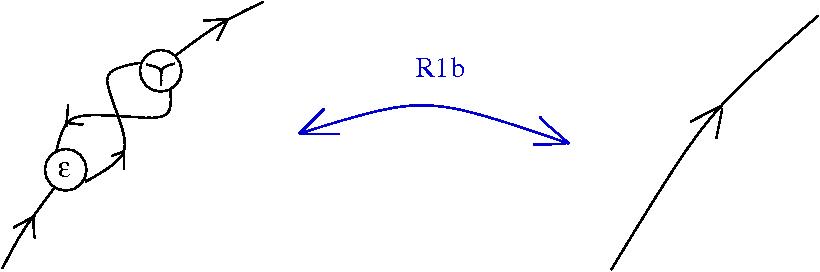}}  \vspace{.5cm}

\paragraph{2.4. (R2) move.} This corresponds to the Reidemeister II move for emergent algebras. It involves an $\Upsilon$ graph and two other: a $\bar{\varepsilon}$ and a $\bar{\mu}$ graph, with $\varepsilon, \mu \in \Gamma$.

\vspace{.5cm}   \centerline{\includegraphics[width=100mm]{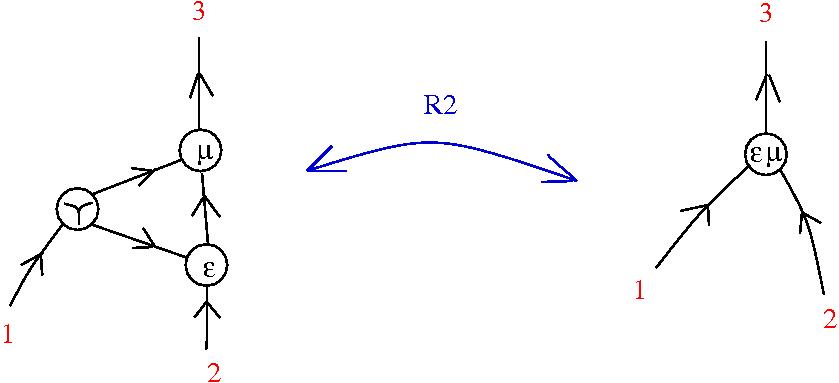}}  \vspace{.5cm}  

This move appears in section 3.4, p. 21 \cite{buligachora}, with the supplementary name "triangle move".

\paragraph{2.5. (ext2) move.} This corresponds to the rule (ext2) from $\lambda$-Scale calculus, it expresses the fact that in emergent algebras the operation indexed with the neutral element $1$ of the group $\Gamma$ has the property $\displaystyle x \circ_{1} y = y$.

\vspace{.5cm}   \centerline{\includegraphics[width=100mm]{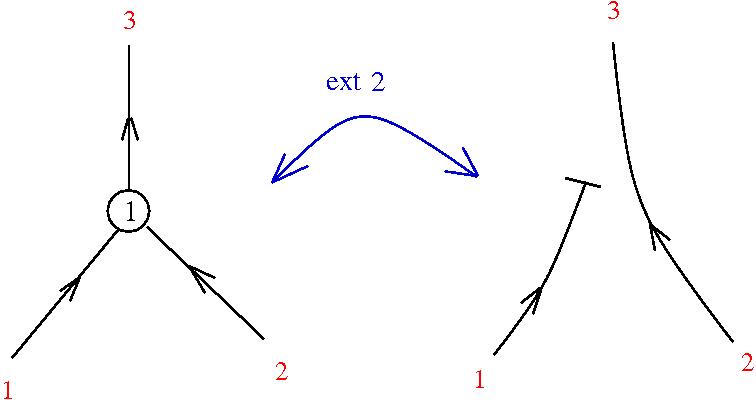}}  \vspace{.5cm} 

\paragraph{2.6. Local pruning.} Local pruning moves are  local moves which eliminate "dead" edges. Notice that, unlike the previous moves, these are one-way (you can eliminate dead edges, but not add them to graphs). 

\vspace{.5cm}   \centerline{\includegraphics[width=80mm]{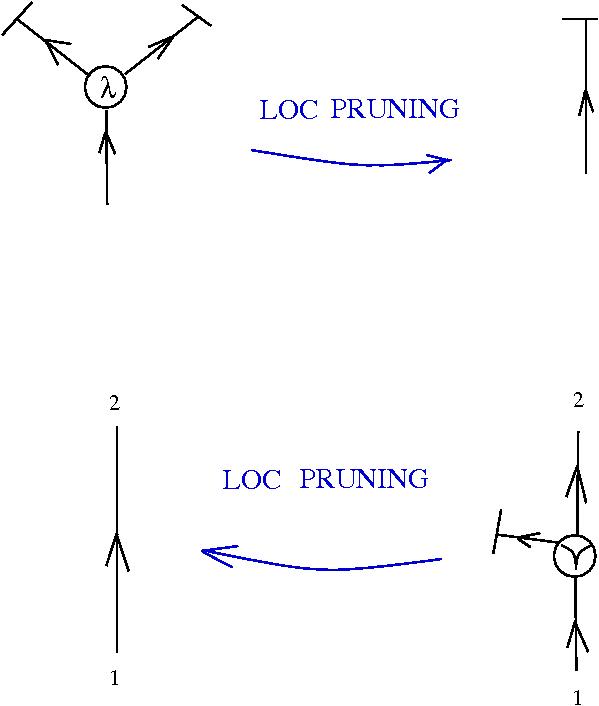}}  \vspace{.5cm}   

\vspace{.5cm}   \centerline{\includegraphics[width=100mm]{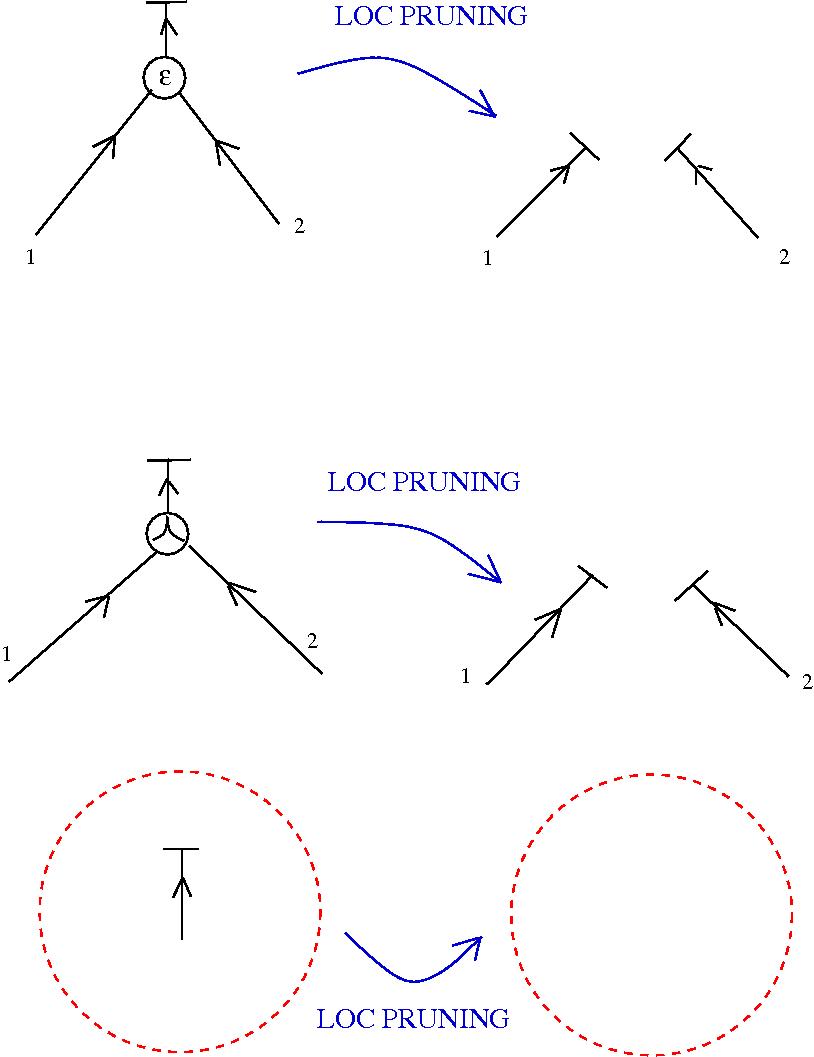}}  \vspace{.5cm}

\paragraph{Global  moves or conditions.} Global  moves are those which are not local, either because the condition $C$ applies to parts of the graph which may have an arbitrary large sum or edges plus nodes, or because after the move the graph $P'$ which replaces the graph $P$ has an arbitrary large sum or edges plus nodes.

\paragraph{2.7. (ext1) move.} This corresponds to the rule (ext1) from $\lambda$-Scale calculus, or 
to $\eta$-reduction in lambda calculus (see  theorem \ref{lambdathm}, part (e) for details). It involves a $\lambda$ graph (think about the $\lambda$ abstraction operation in lambda calculus) and a $\curlywedge$ graph (think about   the application operation in lambda calculus). 

The rule is: if there is no oriented path from "2" to "1", then 
the following move can be performed. 
 
\vspace{.5cm}   \centerline{\includegraphics[width=80mm]{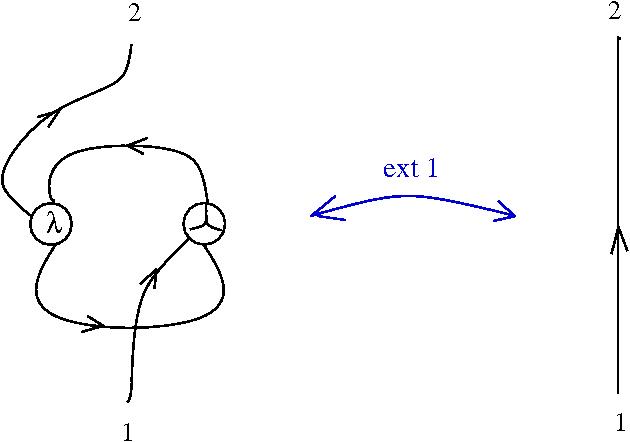}}  \vspace{.5cm} 

\paragraph{2.8. (Global FAN-OUT) move.} This is a global move, because it consists in replacing (under certain circumstances)  a graph by two copies of that graph. 

The rule is: if a graph in $G \in GRAPH$ has a $\Upsilon$ bottleneck, that is if we can find a sub-graph $A \in GRAPH$ connected to the rest of the graph $G$ only through a $\Upsilon$ gate, then we can perform the move  explained in the next figure, from the left to the right.

 Conversely, if in the graph $G$ we can find two identical subgraphs (denoted by $A$), which are in $GRAPH$, which have no edge connecting one with another and which are connected to the rest of $G$ only through one edge, as in the RHS of the figure, then we can perform the move from the right to the left. 

\vspace{.5cm}   \centerline{\includegraphics[width=100mm]{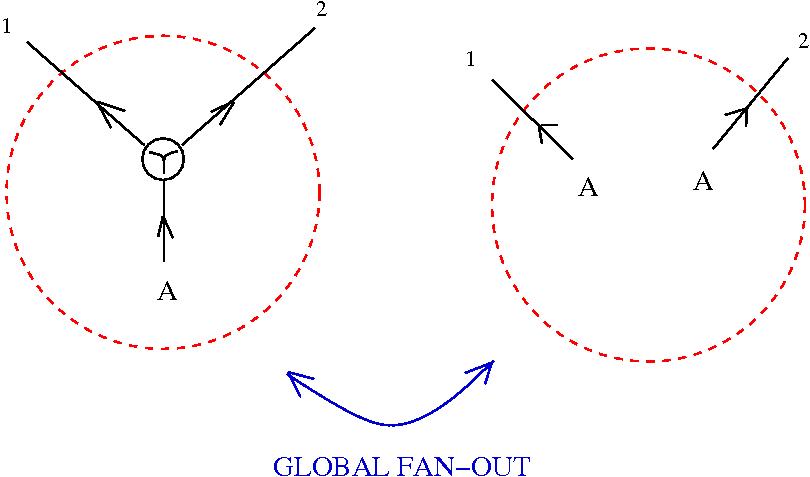}}  \vspace{.5cm}   

Remark that (global FAN-OUT) trivially implies (CO-COMM). ( As an local rule alternative to the global FAN-OUT, we might consider the following.   Fix a number $ N$ and consider only graphs $ A$ which have at most $ N$ (nodes + arrows). The $ N$ LOCAL FAN-OUT move is the same as the GLOBAL FAN-OUT move, only it applies only to such graphs $ A$.   This local FAN-OUT move does not imply CO-COMM.)

\paragraph{2.9. Global pruning.} This a global move which eliminates "dead" edges.

The rule is: if a graph in $G \in GRAPH$ has a $\top$ ending, that is if we can find a sub-graph $A \in GRAPH$  connected only to a  $\top$ gate, with no edges connecting to the rest of $G$, then we can erase this graph and the respective $\top$ gate.

\vspace{.5cm}   \centerline{\includegraphics[width=100mm]{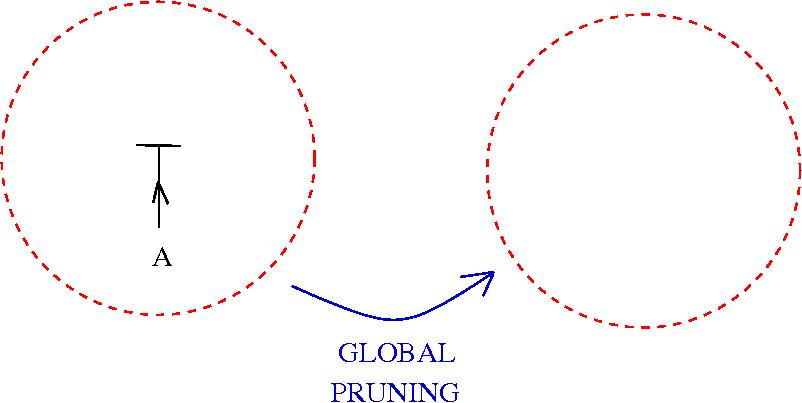}}  \vspace{.5cm}   
   
The global pruning may be needed because of the $\lambda$ gates, which cannot be removed only by local pruning. 

\paragraph{2.10. Elimination of loops.} It is possible that, after using a local or global move, we obtain a graph with an arrow which closes itself, without being connected to any node. 
Here is an example, concerning the application of the graphic $\beta$ move. 
We may erase any such loop, or add one.

\paragraph{$\lambda$GRAPHS.} The edges of an elementary graph $\lambda$ can be numbered unambiguously, clockwise, by 1, 2, 3, such that 1 is the number of the entrant edge.

\begin{definition}
A graph $G \in GRAPH$ is a $\lambda$-graph, notation $G \in \lambda GRAPH$, if: 
\begin{enumerate}
\item[-] it does not have $\bar{\varepsilon}$ gates, 
\item[-] for any node $\lambda$ any oriented path in $G$ starting at the edge 2 of this node can be completed to a path which either terminates in a graph $\top$, or else terminates at the edge 1 of this node. 
\end{enumerate}
\end{definition}
The condition $G \in \lambda GRAPH$ is global, in the sense that  in order to decide if $G \in \lambda GRAPH$ we have to examine parts of the graph which may have an arbitrary large sum or edges plus nodes.

\section{Conversion of lambda terms into $GRAPH$}
\label{constru}

Here I show how to associate to a lambda term a graph in $GRAPH$, then I use this to show that $\beta$-reduction in lambda calculus transforms into the $\beta$ rule for $GRAPH$. (Thanks to Morita Yasuaki for some corrections.)

Indeed, to any term $A \in T(X)$ (where $T(X)$ is the set of lambda terms over the variable 
set $X$) we associate its syntactic tree. The syntactic tree of any lambda term is constructed by using two gates, one corresponding to the $\lambda$ abstraction and the other corresponding to the application. We draw syntactic trees with the leaves (elements of $X$) at the bottom and the root at the top. We shall use the following notation for the two gates: at the left is the gate for the $\lambda$ abstraction and at the right is the gate for the application. 

\vspace{.5cm}

\centerline{\includegraphics[width=60mm]{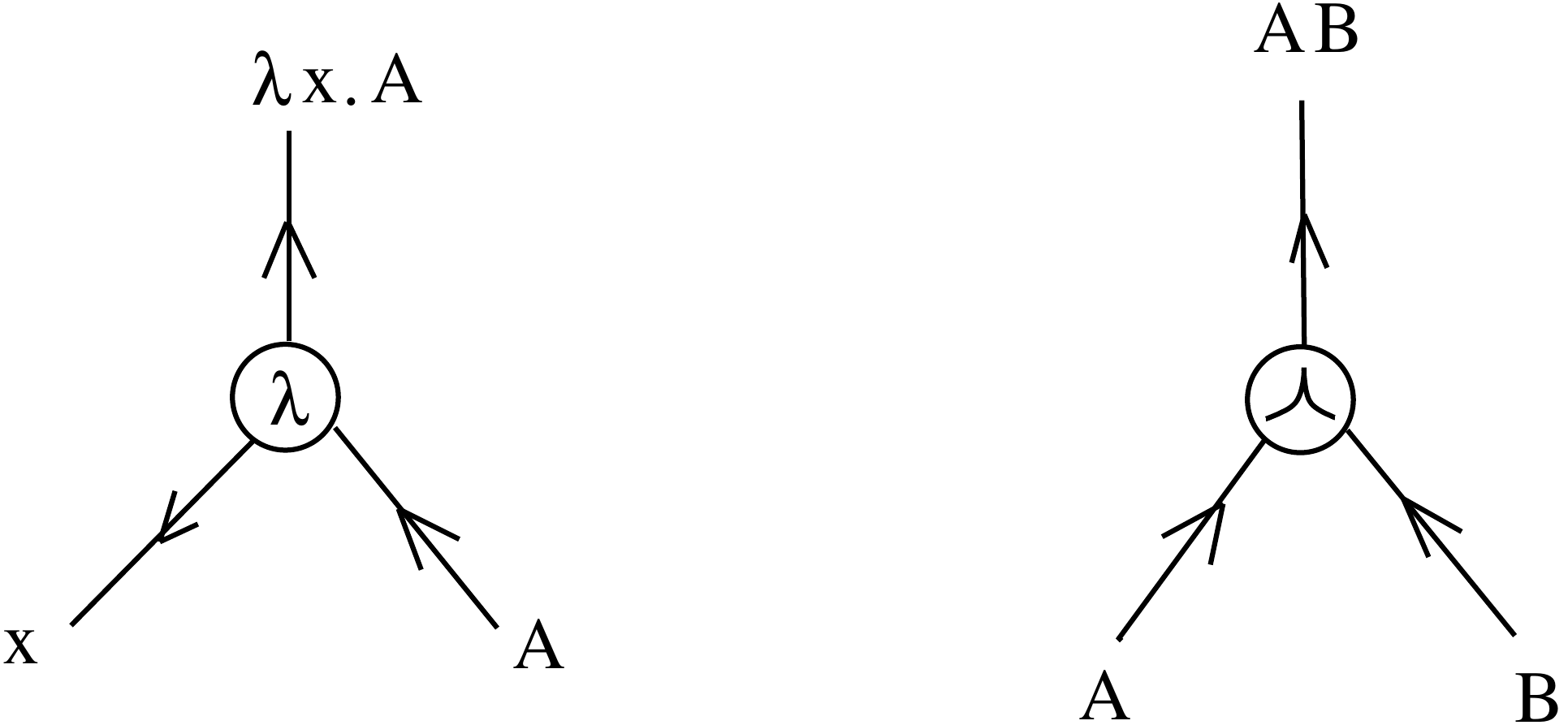}}

\vspace{.5cm}

Remark that these two gates are from the graphical alphabet of $GRAPH$, but the syntactic tree is decorated: at the bottom we have leaves from $X$. Also, remark the peculiar orientation of the edge from the left (in tree notation convention) of the $\lambda$ gate. For the moment, this orientation is in contradiction with the implicit orientation (from down-up) of edges of the syntactic tree, but soon this matter will become clear.

We shall remove all leaves decorations, with the price of introducing new gates, namely $\Upsilon$ and $\top$ gates. This will be done in a sequence of steps, detailed further. Take the syntactic tree of $A \in T(X)$, drawn with the mentioned conventions (concerning gates and the positioning of leaves and root respectively). 

We take as examples the following five lambda terms: $\displaystyle I = \lambda x . x$, 
$\displaystyle  K = \lambda x . (\lambda y. x)$, $\displaystyle S = \lambda x . ( \lambda y . 
(\lambda z . ((xz)(yz))))$, $\displaystyle \Omega = (\lambda x. (xx)) (\lambda x. (xx))$ and 
$\displaystyle T =  (\lambda x. (xy)) (\lambda x. (xy))$. 

\paragraph{Step 1.} Elimination of bound variables, part I. Any leaf of the tree is connected to the root by an unique path. 

Start from the leftmost leaf, perform the algorithm explained further, then   go to the right and repeat until all leaves are exhausted. We initialize also a list $B  = \emptyset$ of bound variables.

Take a leaf, say decorated with $x \in X$. To this leaf is associated a word (a list) which is formed by the symbols of gates which are on the path which connects (from the bottom-up) the leaf with the root, together with information about which way, left (L) or right (R), the path passes through the gates. Such a word is formed by the letters $\displaystyle \lambda^{L}$,  
$\displaystyle \lambda^{R}$,  $\displaystyle \curlywedge^{L}$, $\displaystyle \curlywedge^{R}$.

If the first letter is $\displaystyle \lambda^{L}$ then add to the list  $B$ the pair 
$(x, w(x))$ formed by the variable name $x$, and the associated word (describing the path to follow from the respective leaf to the root). Then pass to a new leaf. 

Else continue along the path to the roof. If we arrive at a $\lambda$ gate, this can happen only coming from the right leg of the $\lambda$ gate, thus we can find only the letter $\displaystyle \lambda^{R}$. In such a case look at the variable $y$ which decorates the left leg of the same $\lambda$ gate. If $x =y$ then add to the syntactic tree a new edge, from $y$ to $x$ and proceed further along the path, else proceed further. If the root is attained then pass to next leaf.  

Examples: the graphs associated to the mentioned lambda terms,  together with the list of bound variables, are the following. 

\begin{enumerate}
\item[-] $\displaystyle I = \lambda x . x$ has $\displaystyle B = \left\{ (x, \lambda^{L}) \right\}$, $\displaystyle  K = \lambda x . (\lambda y. x)$ has $\displaystyle B = \left\{ (x,  \lambda^{L}), (y, \lambda^{L} \lambda^{R})  \right\}$, $\displaystyle S = \lambda x . ( \lambda y . (\lambda z . ((xz)(yz))))$ has $\displaystyle B = \left\{ (x,  \lambda^{L}), (y, \lambda^{L} \lambda^{R}), (z, \lambda^{L} \lambda^{R} \lambda^{R})  \right\}$. 
\vspace{.5cm}

\centerline{\includegraphics[width=90mm]{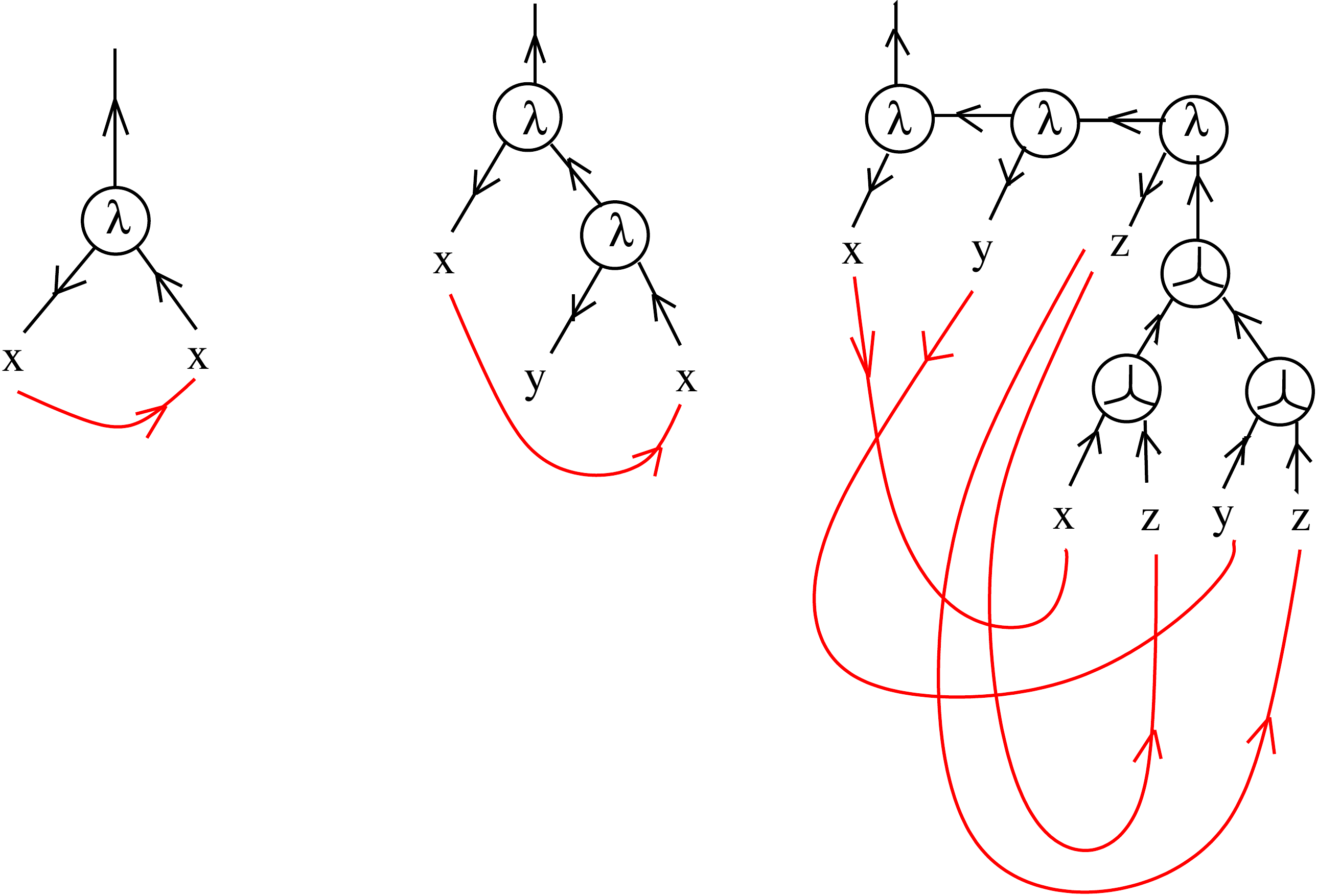}}

\vspace{.5cm}

\item[-]  $\displaystyle \Omega = (\lambda x. (xx)) (\lambda x. (xx))$ has $\displaystyle B = \left\{ (x, \lambda^{L} \curlywedge^{L}) , (x, \lambda^{L} \curlywedge^{R}) \right\}$, 
$\displaystyle T =  (\lambda x. (xy)) (\lambda x. (xy))$ has $\displaystyle B = \left\{ (x, \lambda^{L} \curlywedge^{L}) , (x, \lambda^{L} \curlywedge^{R}) \right\}$. 
\vspace{.5cm}

\centerline{\includegraphics[width=90mm]{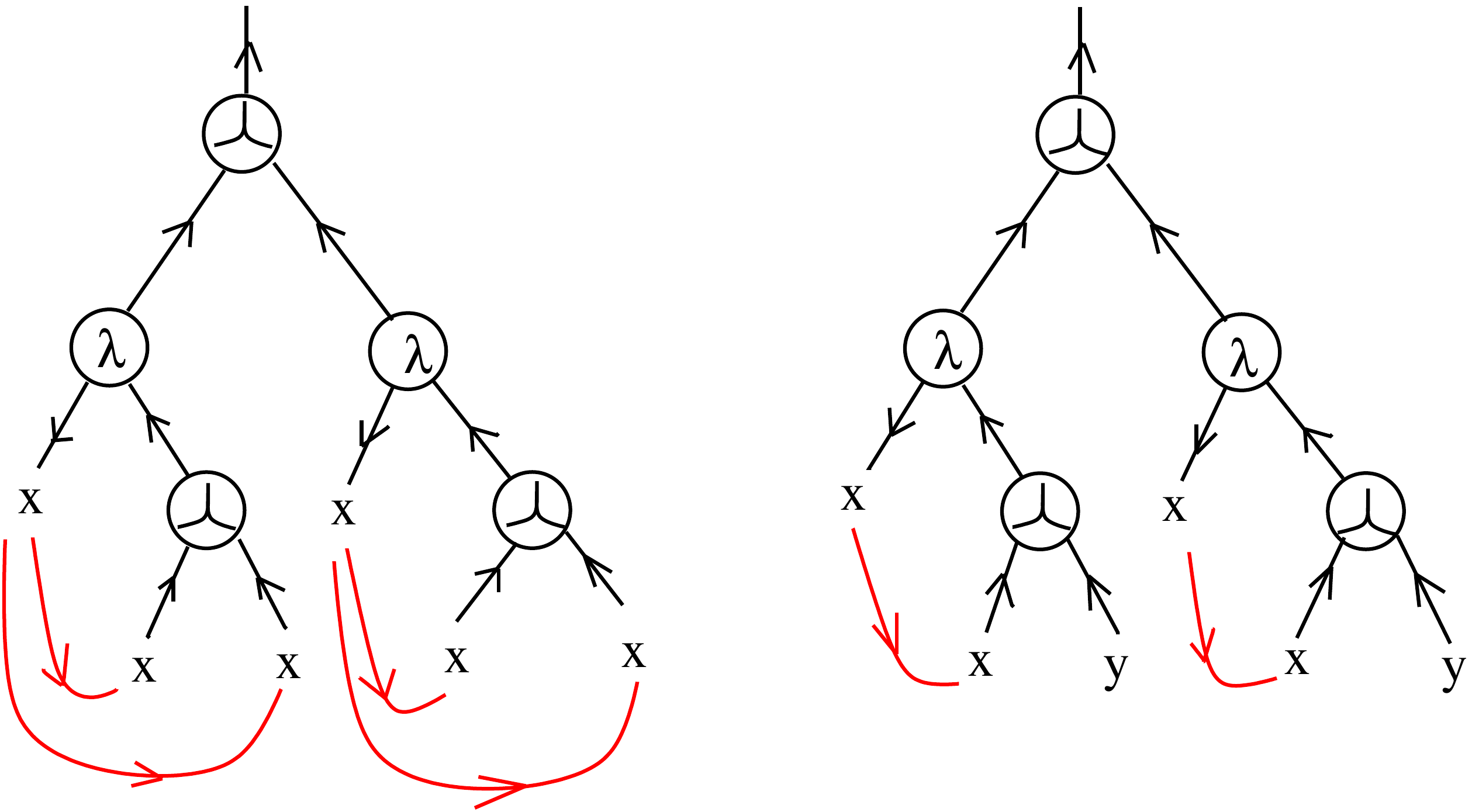}}

\vspace{.5cm}

\end{enumerate}

\paragraph{Step 2.} Elimination of bound variables, part II. We have now a list $B$ of bound variables. If the list is empty then go to the next step. Else, do the following, starting from the first element of the list, until the list is finished. 

An element, say $(x, w(x))$, of the list, is either connected to other leaves by one or more edges added at step 1, or not. If is not connected then erase the variable name with the associated path $w(x)$  and replace it by a $\top$ gate. 
If it is connected then erase it, replace it by a tree formed by $\Upsilon$ gates, which starts at the place where the element of the list were before the erasure and stops at the leaves which were connected to $x$. Erase all decorations which were joined to $x$ and also erase all edges which were added at step 1 to the leave $x$ from the list.

Examples: after the step 2, the graphs associated to the mentioned lambda terms  are the following. 

\begin{enumerate}
\item[-] the graphs of  $\displaystyle I = \lambda x . x$, $\displaystyle  K = \lambda x . (\lambda y. x)$, $\displaystyle S = \lambda x . ( \lambda y . (\lambda z . ((xz)(yz))))$ are  
\vspace{.5cm}

\centerline{\includegraphics[width=90mm]{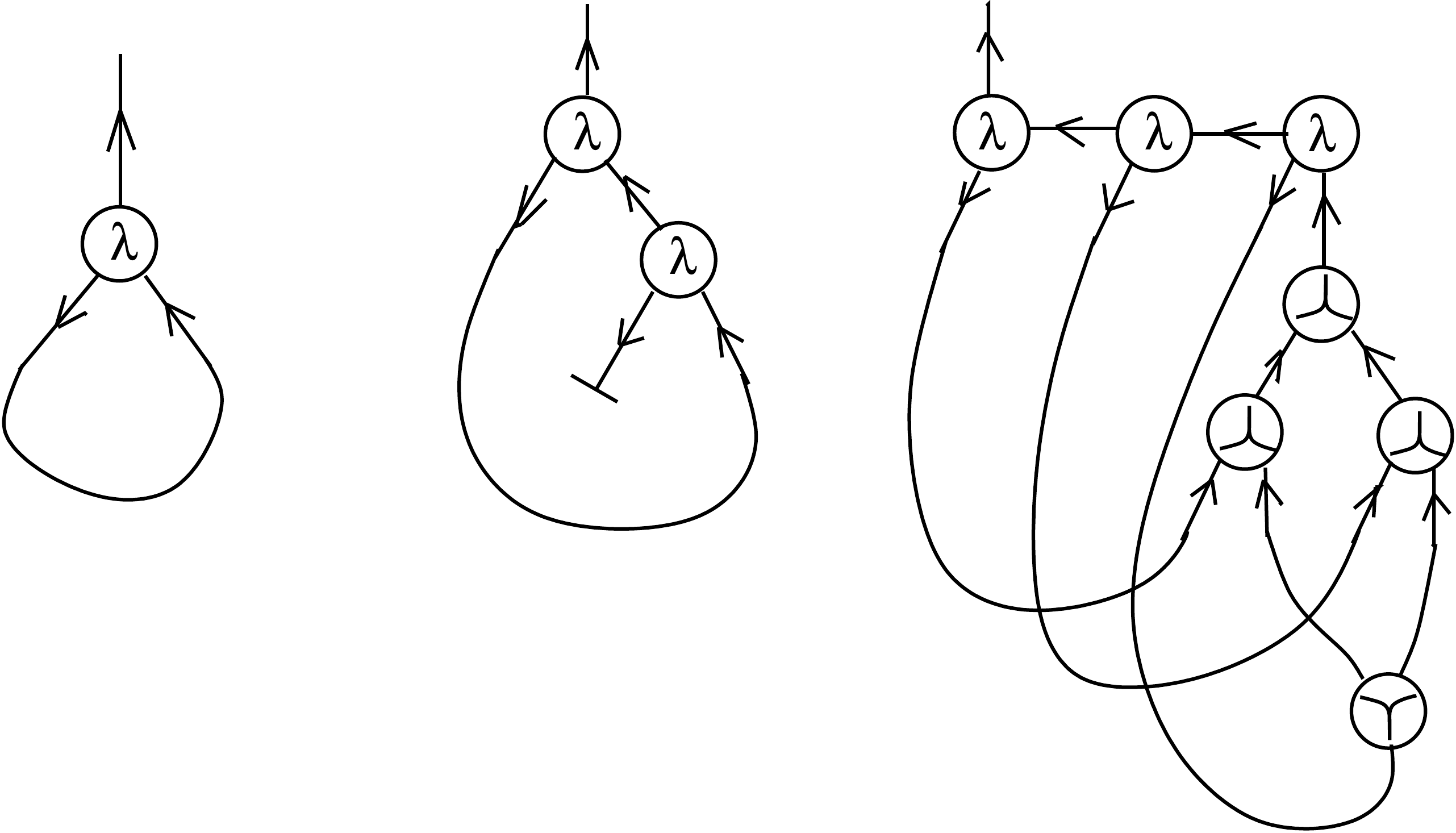}}

\vspace{.5cm}

\item[-]  the graphs of $\displaystyle \Omega = (\lambda x. (xx)) (\lambda x. (xx))$,  
$\displaystyle T =  (\lambda x. (xy)) (\lambda x. (xy))$ are  
\vspace{.5cm}

\centerline{\includegraphics[width=90mm]{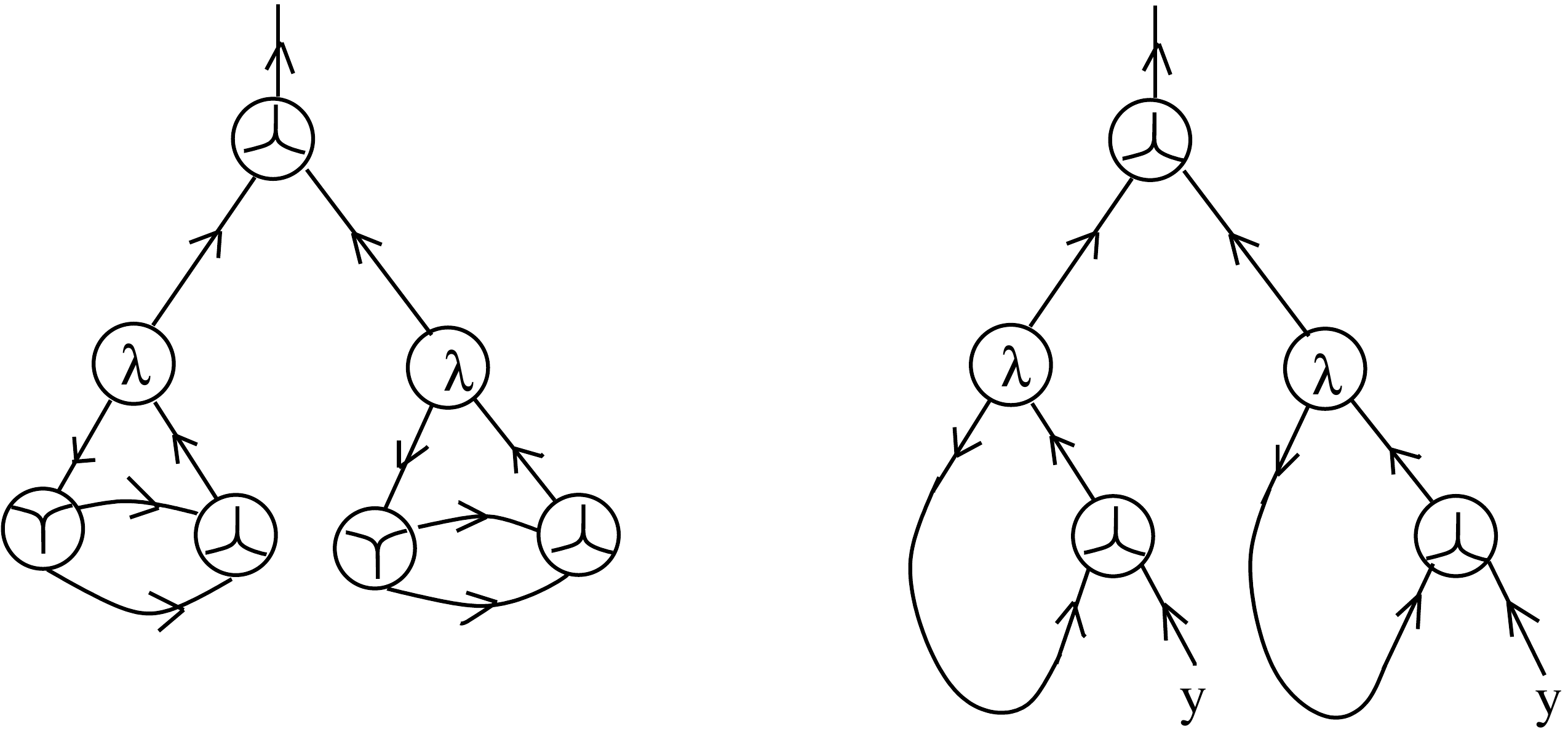}}

\vspace{.5cm}

\end{enumerate}

Remark that at this step the necessity of having the peculiar orientation of the left leg of the $\lambda$ gate becomes clear. 

Remark also that there may be more than one possible tree of gates $\Upsilon$, at each elimination of a bound variable (in case a bound variable has at least tree occurrences). One may use any tree of $\Upsilon$ which is fit. The problem of multiple possibilities is the reason of introducing the (CO-ASSOC) move.  

\paragraph{Step 3.} We may still have leaves decorated by free variables. Starting from the left to the right, group them together in case some of them occur in multiple places, then replace the multiple occurrences of a free variable by a tree of $\Upsilon$ gates with a free  root, which ends exactly where the occurrences of the respective variable are. Again, there are multiple ways of doing this, but we may pass from one to another by a sequence of (CO-ASSOC) moves. 

Examples: after the step 3,  all the graphs associated to the mentioned lambda terms, excepting the last one, are left unchanged. The graph of the last term, changes. 

\begin{enumerate}
\item[-]  as an illustration, I figure the graphs of $\displaystyle \Omega = (\lambda x. (xx)) (\lambda x. (xx))$,  left unchanged by step 3, and the graph of 
$\displaystyle T =  (\lambda x. (xy)) (\lambda x. (xy))$:
\vspace{.5cm}

\centerline{\includegraphics[width=90mm]{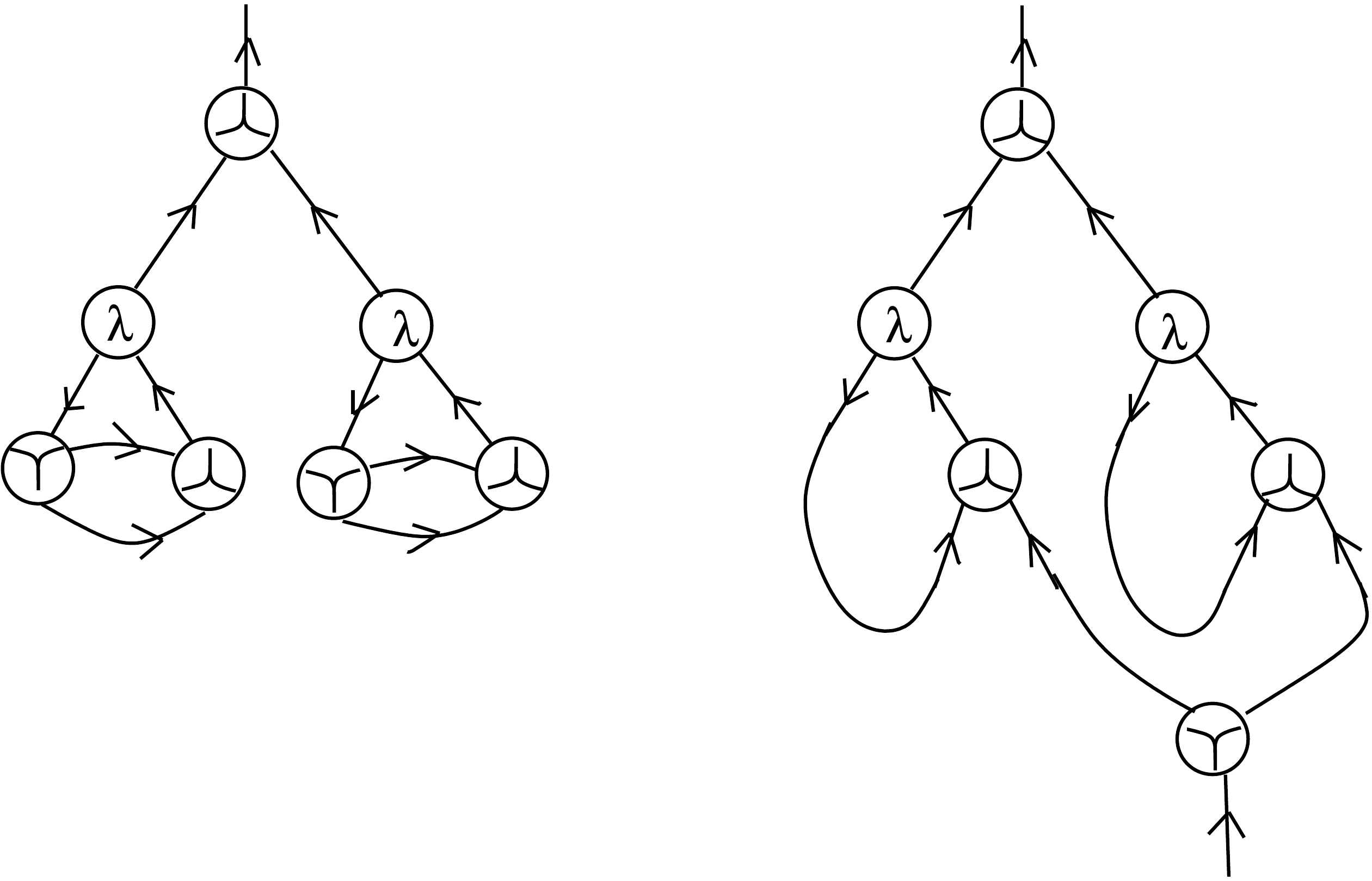}}

\vspace{.5cm}

\end{enumerate}

\begin{theorem}
Let $A \mapsto [A]$ be a transformation of a lambda term $A$ into a graph $[A]$ as described previously (multiple transformations are possible because of the choice of $\Upsilon$ trees). Then: 
\begin{enumerate}
\item[(a)] for any term $A$ the graph $[A]$ is in $\lambda GRAPH$, 
\item[(b)] if $[A]'$ and $[A]"$ are transformations of the term $A$ then we may pass from 
$[A]'$ to $[A]"$ by using a finite number (exponential in the number of leaves of the syntactic tree of $A$) of (CO-ASSOC) moves,
\item[(c)] if $B$ is obtained from $A$ by $\alpha$-conversion then we may pass from  $[A]$ to $[B]$ by a finite sequence of (CO-ASSOC) moves, 
\item[(d)] let $A, B \in T(X)$ be two terms and $x \in X$ be a variable. Consider the terms 
$\lambda x . A$ and $A[x:=B]$, where $A[x:=B]$ is the term obtained by substituting in $A$ the free occurrences of $x$ by $B$. We know that $\beta$ reduction in lambda calculus consists in passing from $(\lambda x . A) B$ to $A[x:=B]$.  Then, by one $\beta$ move in $GRAPH$ applied to  $[(\lambda x . A) B]$ we pass to a graph which can be further  transformed into one  of  $A[x:=B]$, via (global FAN-OUT) moves,  (CO-ASSOC) moves and pruning moves,
\item[(e)] with the notations from (d), consider the terms $A$ and $\lambda x . Ax$ with $x \not \in FV(A)$; then the $\eta$ reduction, consisting in passing from $\lambda x. Ax$ to $A$, corresponds to the ext1 move applied to the graphs $[\lambda x. Ax]$ and $[A]$. 
\end{enumerate}
\label{lambdathm}
\end{theorem} 

\paragraph{Proof.} (a) we have to prove that for any node $\lambda$ any oriented path in $[A]$ starting at the left exiting edge  of this node can be completed to a path which either terminates in a graph $\top$, or else terminates at the entry peg of this node, but this is clear. Indeed, either the bound variable (of this $\lambda$ node in the syntactic tree of $A$) is fresh, then the bound variable is replaced by a $\top$ gate, or else, the bound variable is replaced by a tree of $\Upsilon$ gates. No matter which path we choose, we may complete it to a cycle passing by the said $\lambda$ node.

(b) Clear also, because the (CO-ASSOC) move is designed for passing from a tree of $\Upsilon$ gates to another tree with the same number of leaves. 

(c) Indeed, the names of bound variables of $A$ do not affect the construction of $[A]$, therefore if $B$ is obtained by $\alpha$-conversion of $A$, then $[B]$ differs from $[A]$ only by the particular choice of  trees of $\Upsilon$ gates. But this is solved by (CO-ASSOC) moves. 

(d) This is the surprising, maybe, part of the theorem. There are two cases: $x$ is fresh for $A$ or not. If $x$ is fresh for $A$ then in the graph $[(\lambda x.A) B]$ the name variable $x$ is replaced by a $\top$ gate. If not, then all the occurrences of $x$ in $A$ are connected by a $\Upsilon$ tree with root at the left peg of the $\lambda$ gate where $x$ appears as a bound variable.

In the case when $x$ is not fresh for $A$, we see in the LHS of the figure the graph  $[(\lambda x . A) B]$  (with a remanent decoration of "x").  We perform a graphic ($\beta$) move and we obtain the graph from the right.

\vspace{.5cm}

\centerline{\includegraphics[width=90mm]{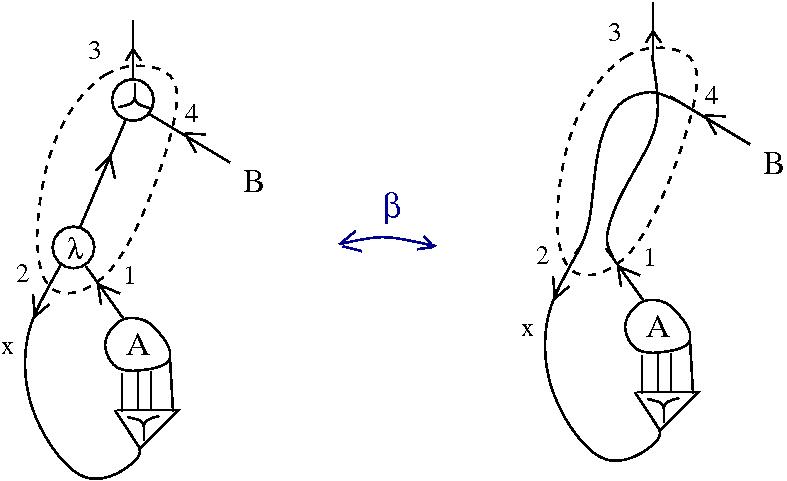}}

\vspace{.5cm}

This graph can be transformed into a graph of $A[x:=B]$ via (global FAN-OUT) and (CO-ASSOC)  moves. 
The case when  $x$ is fresh for $A$ is figured next.

\vspace{.5cm}

\centerline{\includegraphics[width=90mm]{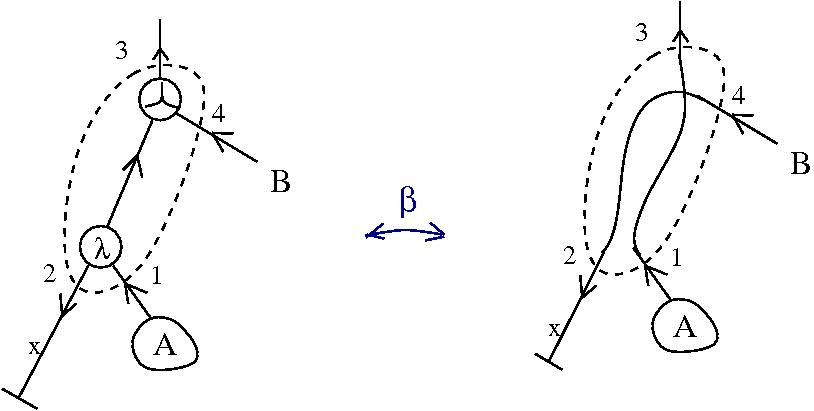}}

\vspace{.5cm}

We see that the graph obtained by performing the graphic ($\beta$) move is the union of the graph of $A$ and the graph of $B$ with a $\top$ gate added at the root. By pruning we are left with the graph of $A$, which is consistent to the fact that when  $x$ is fresh for $A$ then 
$(\lambda x . A) B$ transforms by  $\beta$ reduction into $A$. 

(e) In the next figure we see at the LHS the graph $[\lambda x . Ax]$ and at the RHS the graph $[A]$. 

\vspace{.5cm}

\centerline{\includegraphics[width=80mm]{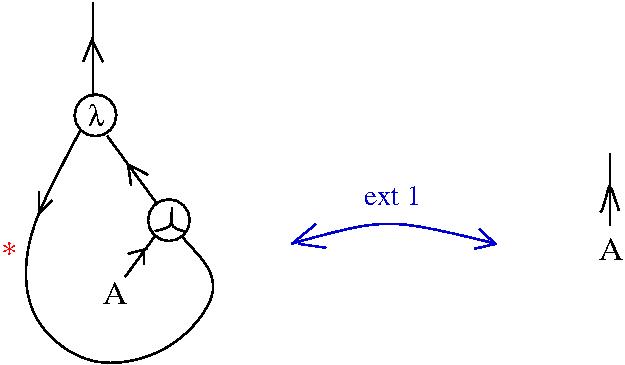}}

\vspace{.5cm}

The red asterisk marks the arrow which appears in the construction $[\lambda x . Ax]$ from  the variable $x$, taking into account the hypothesis $x \not \in FV(A)$. We have a pattern where we can apply the ext1  move and we obtain $[A]$, as claimed. \quad $\square$

As an example, let us manipulate the graph of $\displaystyle \Omega = (\lambda x. (xx)) (\lambda x. (xx))$: 

\vspace{.5cm}

\centerline{\includegraphics[width=90mm]{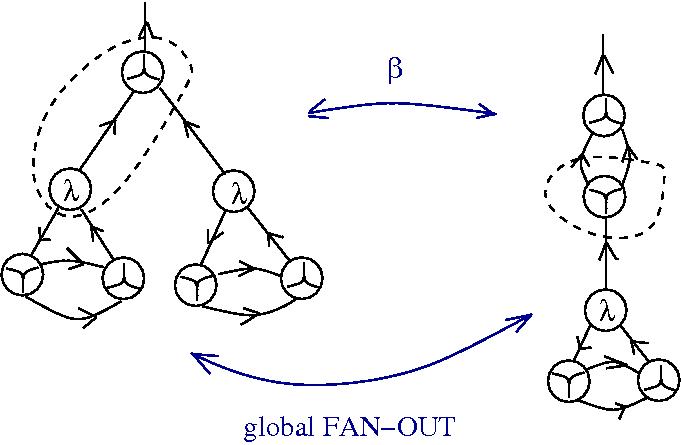}}

\vspace{.5cm}

We can pass from the LHS figure to the RHS figure by using a graphic ($\beta$) move. Conversely, we can pass from the RHS figure to the LHS figure by using a (global FAN-OUT) move. 
These manipulations correspond to the well known fact that $\Omega$ is left unchanged after $\beta$ reduction:  let $\displaystyle U = \lambda x. (xx)$, then 
$\displaystyle \Omega = U U = (\lambda x. (xx)) U \leftrightarrow  U U = \Omega$.

\subsection{Example: combinatory logic}

\paragraph{$S$, $K$ and $I$ combinators in $GRAPH$.} The combinators $\displaystyle I = \lambda x . x$, $\displaystyle  K = \lambda x . (\lambda y. (xy))$ and $\displaystyle S = \lambda x . ( \lambda y . (\lambda z . ((xz)(yz))))$ have the following correspondents in $GRAPH$, denoted by the same letters: 

\vspace{.5cm}

\centerline{\includegraphics[width=100mm]{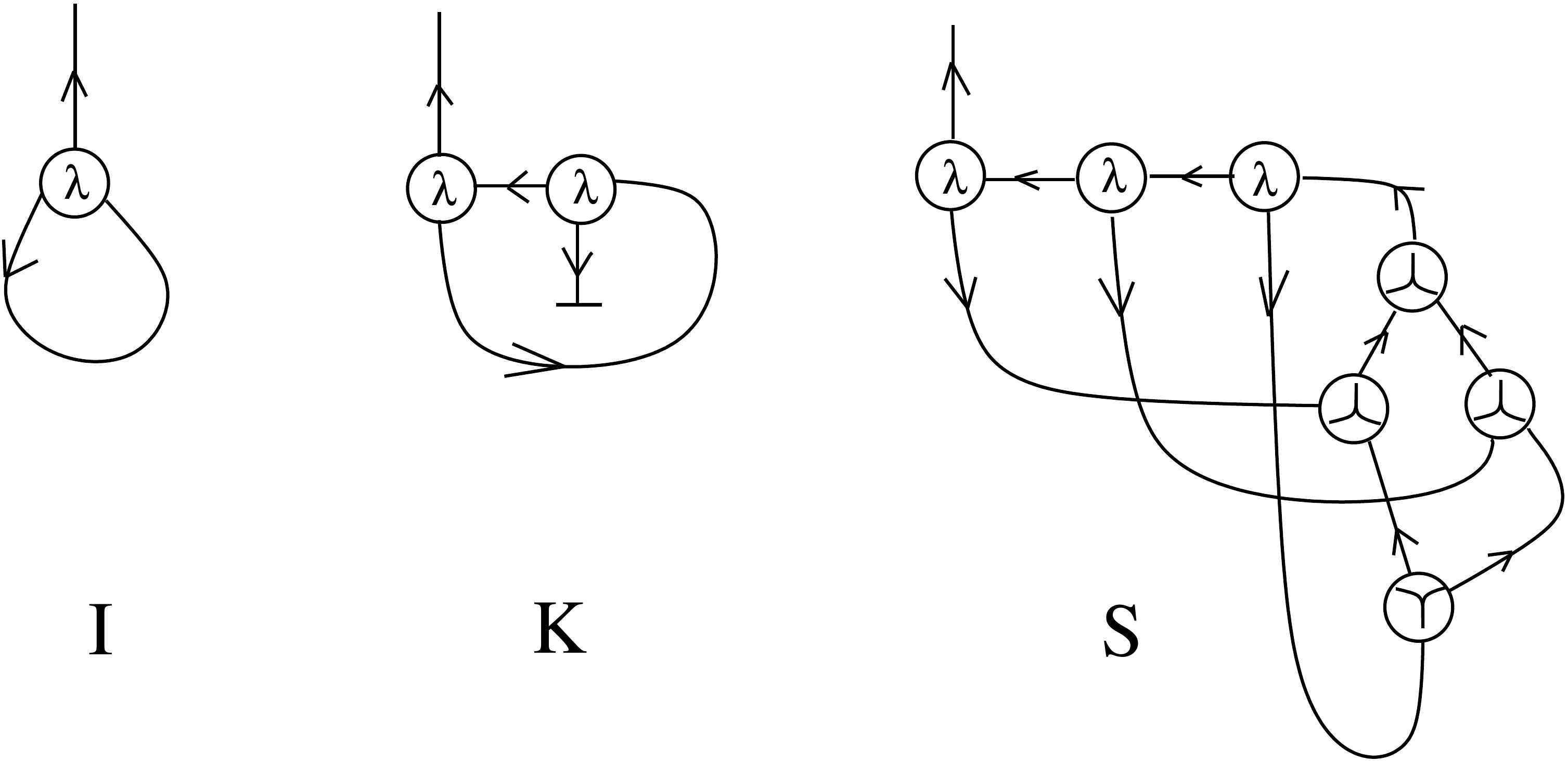}}

\vspace{.5cm}

\begin{proposition}
(a) By one graphic ($\beta$) move $I \curlywedge A$ transforms into $A$, for any $A \in GRAPH$ with one output. 

(b) By two graphic ($\beta$) moves, followed by a global pruning, for any $A, B \in GRAPH$ with one output, the graph $(K \curlywedge A) \curlywedge B$ transforms into $A$. 

(c) By five graphic ($\beta$) moves, followed by one local pruning move, the graph $(S \curlywedge K) \curlywedge K$ transforms into $I$. 

(d) By three graphic ($\beta$) moves followed by a (global FAN-OUT) move, for any  $A, B, C \in GRAPH$ with one output, the graph 
$((S \curlywedge A)\curlywedge B) \curlywedge C$ transforms into the graph $(A \curlywedge C) \curlywedge (B \curlywedge C)$. 
\label{pcombi}
\end{proposition}

\paragraph{Proof.} The proof of (b) is given in the next figure. 

\vspace{.5cm}

\centerline{\includegraphics[width=120mm]{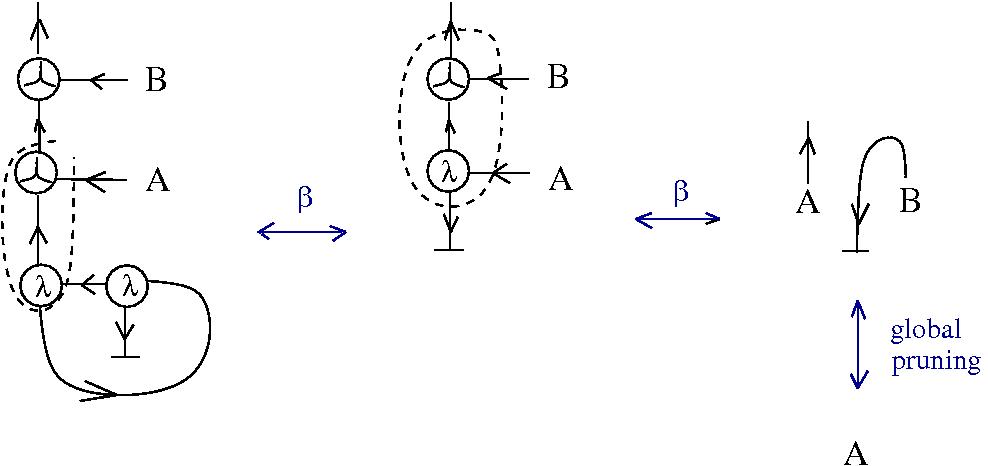}}

\vspace{.5cm}

The proof of (c) is given in the following figure. 

\vspace{.5cm}

\centerline{\includegraphics[width=120mm]{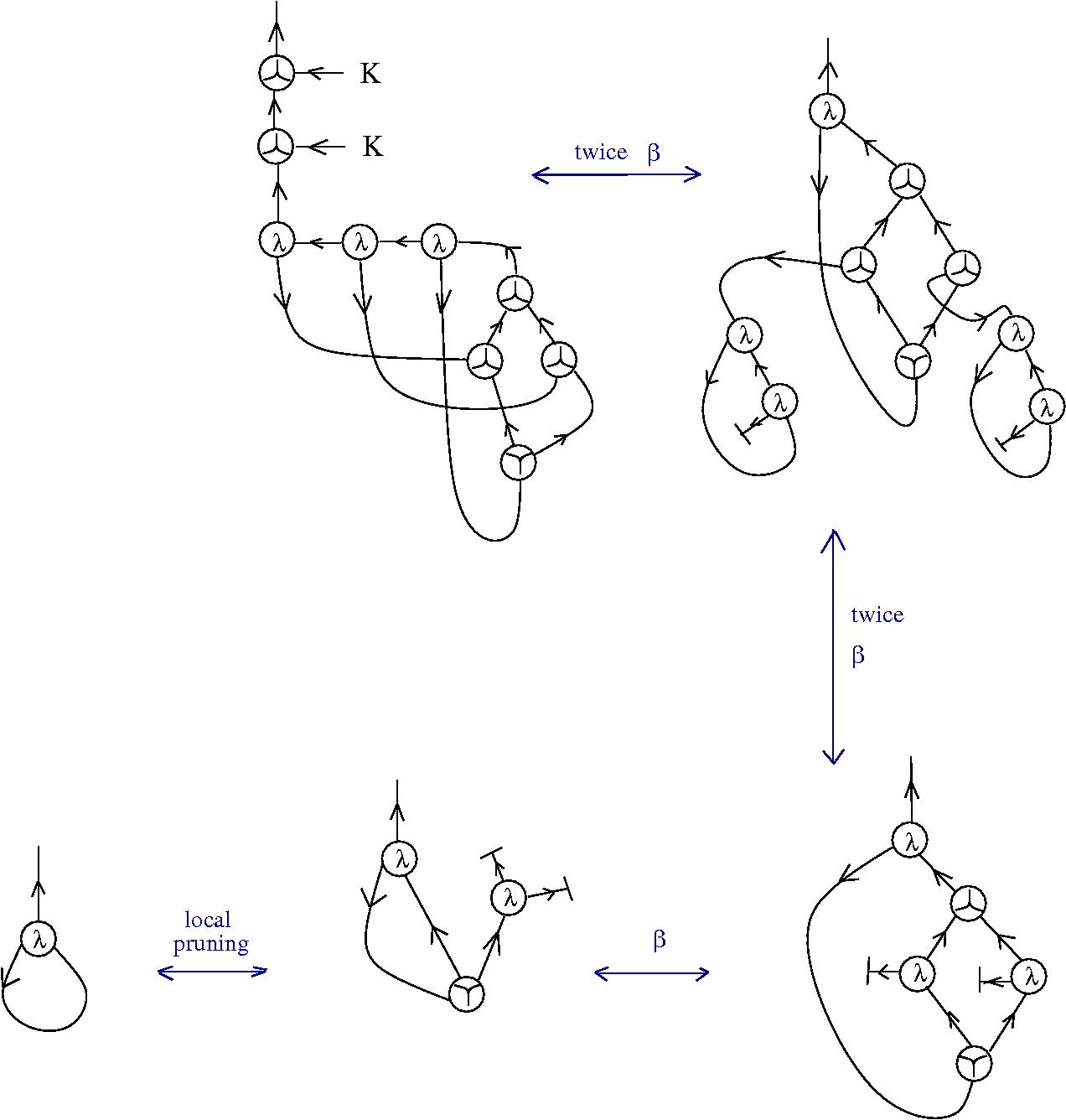}}

\vspace{.5cm}

 (a) and (d) are left to the interested reader. \quad $\square$

\section{Using graphic lambda calculus}
\label{around}

The manipulations of graphs presented in this section can  be applied for graphs which represent lambda terms. However, they can also be applied for graphs which do not represent lambda terms. 

\paragraph{Fixed points.} Let's start with a graph $A \in GRAPH$, which has one distinguished input and one distinguished  output.I represent  this as follows. 

 \vspace{.5cm}

\centerline{ \includegraphics[width=20mm]{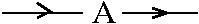}}

\vspace{.5cm}

For any  graph $\displaystyle B$ with one output, we denote by $\displaystyle A(B)$ the graph obtained  by grafting the output of $\displaystyle B$ to the input of $\displaystyle A$.

I want to  find $\displaystyle B$ such that $\displaystyle A(B) \leftrightarrow B$, where $\displaystyle \leftrightarrow$ means any finite sequence of moves in graphic lambda calculus. I call such a graph $B$ a fixed point of $A$. 

The solution of this problem  is the same as in usual lambda calculus.  We start from the following succession of moves:

 \vspace{.5cm}

\centerline{ \includegraphics[width=80mm]{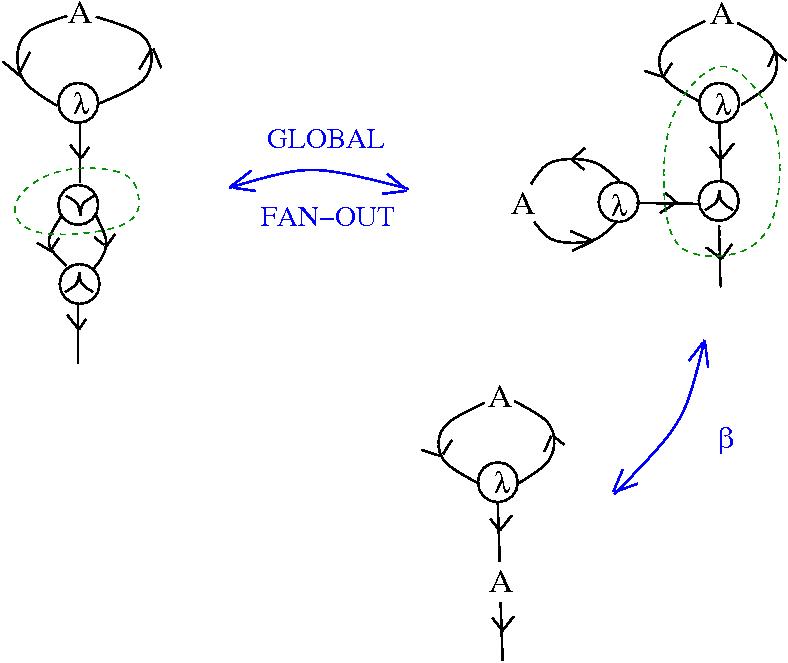}}

\vspace{.5cm}

This is very close to the solution, we only need a small modification:

 \vspace{.5cm}

\centerline{ \includegraphics[width=80mm]{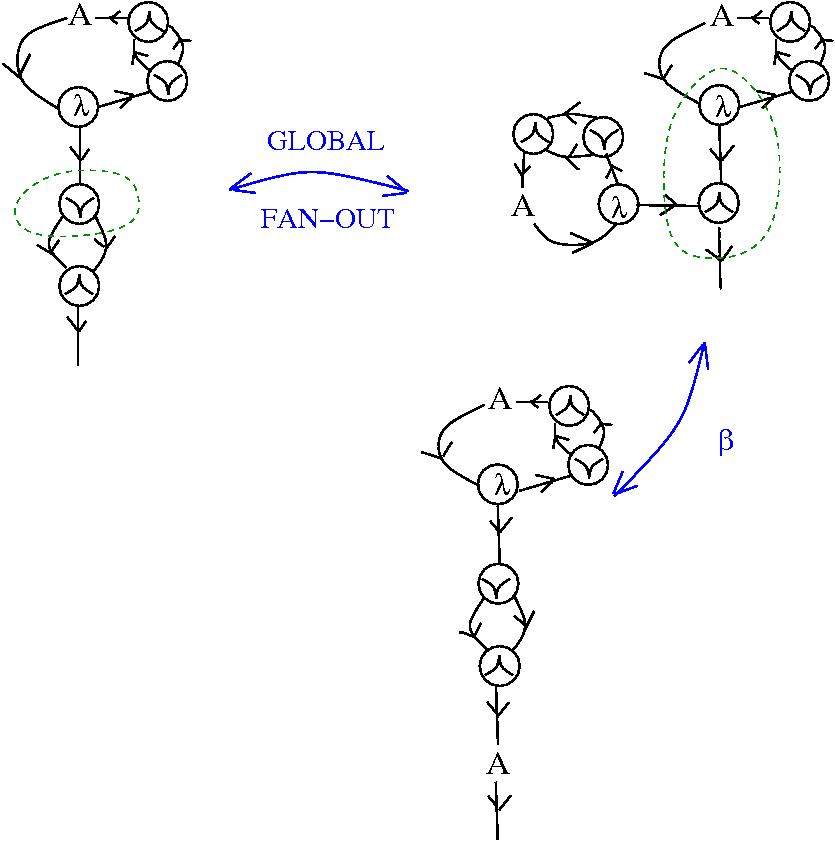}}

\vspace{.5cm}

\paragraph{Grafting, application or abstraction?} If the $A$, $B$ from the previous paragraph were representing lambda terms, then the natural operation between them is not grafting, but the application. Or, in graphic lambda calculus the application it's represented by an elementary graph, therefore $AB$ (seen as the term in lambda calculus which is obtained as the application of $A$ to $B$) is not represented as a grafting of the output of $B$ to the input of $A$. 

We can easily transform grafting into the application operation. 

 \vspace{.5cm}

\centerline{ \includegraphics[width=80mm]{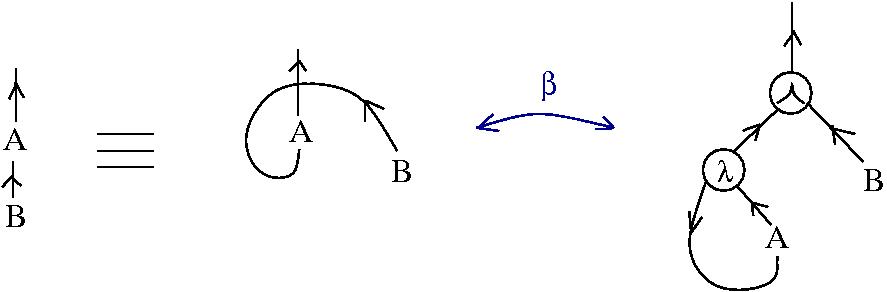}}

\vspace{.5cm}

Suppose that  $A$ and $B$ are graphs representing lambda terms, more precisely suppose that $A$ is representing a term (denoted by $A$ too) and it's input represents a free variable $x$ of the term  $A$. Then the grafting of $B$ to $A$ is the term $A[x:=B]$ and  the graph from the right is representing $(\lambda x.A)B$, therefore both graphs are representing terms from lambda calculus. 

We can transform grafting into something else: 

 \vspace{.5cm}

\centerline{ \includegraphics[width=80mm]{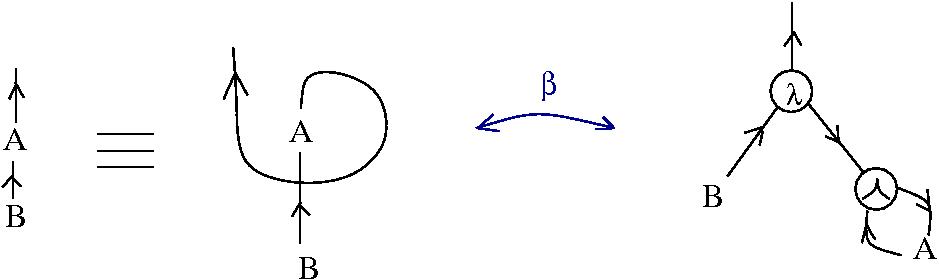}}

\vspace{.5cm}

This has no meaning in lambda calculus, but excepting the orientation of one of the arrows of the graph from the right, it looks  like if the abstraction gate (the $\lambda$ gate) plays the role of an application operation. 

\paragraph{Zippers and combinators as half-zippers.} Let's take $\displaystyle  n \geq 1$ a natural number and let's consider the following graph in $\displaystyle  GRAPH$, called the n-zipper:

 \vspace{.5cm}

\centerline{ \includegraphics[width=65mm]{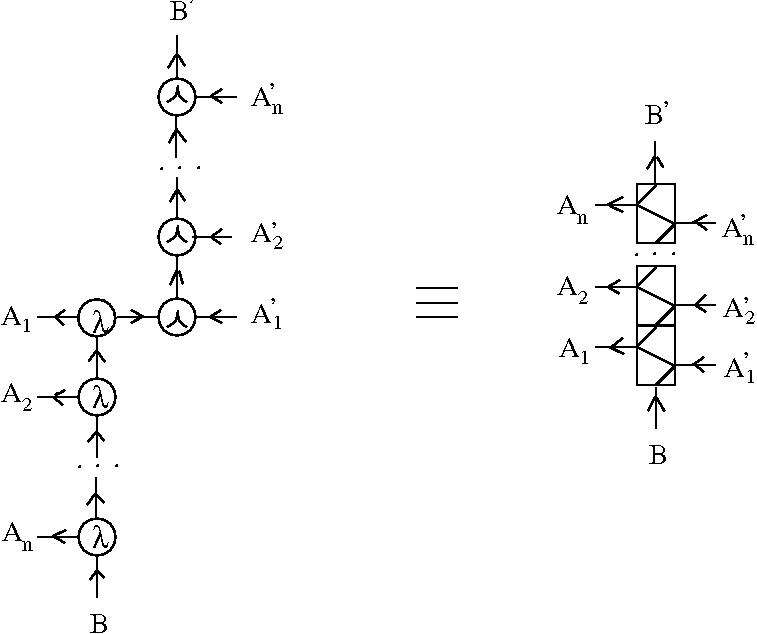}}

\vspace{.5cm}

At the left is the n-zipper graph; at the right is a notation for it, or a "macro".  
The zipper graph is interesting because it allows to perform  (nontrivial) graphic beta moves in a fixed order.   In the following picture is figured in red the place where the first graphic beta move is applied.

 \vspace{.5cm}

\centerline{ \includegraphics[width=65mm]{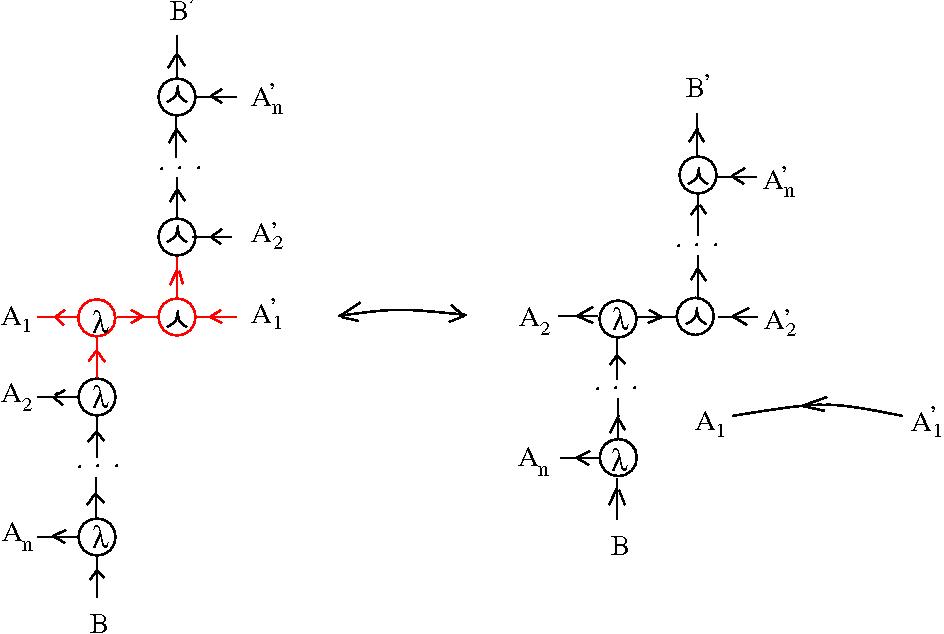}}

\vspace{.5cm}

In terms of zipper notation this graphic beta move has the following appearance:

 \vspace{.5cm}

\centerline{ \includegraphics[width=65mm]{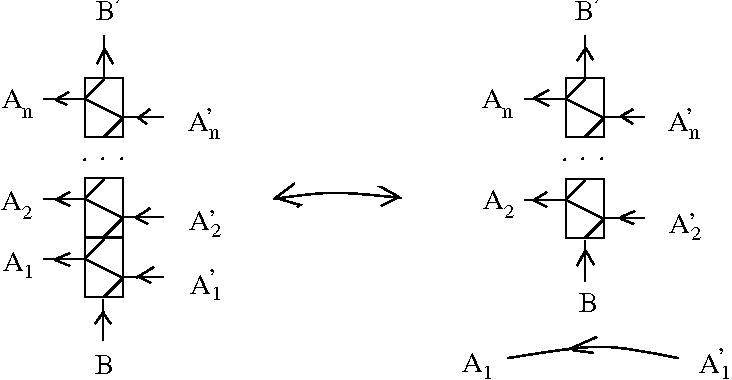}}

\vspace{.5cm}

We see that a n-zipper transforms into a (n-1)-zipper plus an arrow. We may repeat this move, as long as we can. This procedure defines a "zipper move":

 \vspace{.5cm}

\centerline{ \includegraphics[width=65mm]{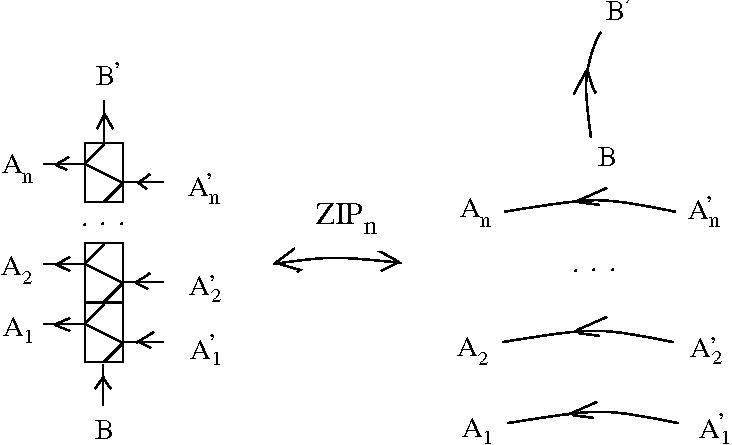}}

\vspace{.5cm}

We may see the 1-zipper move as  the graphic beta move, which transforms the 1-zipper into two arrows.

The combinator $\displaystyle  I = \lambda x . x$  satisfies the relation $\displaystyle  I A = A$. In the next figure it is shown that $\displaystyle  I$  (figured in green), when applied to $A$,  is just a half of the 1-zipper, with an arrow added (figured in blue).

 \vspace{.5cm}

\centerline{ \includegraphics[width=80mm]{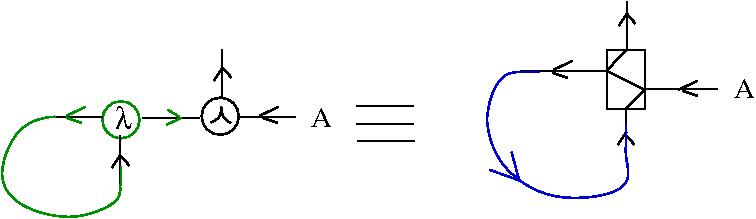}}

\vspace{.5cm}

By opening the zipper we obtain $\displaystyle  A$, as it should.

The combinator $\displaystyle  K = \lambda xy.x$ satisfies $\displaystyle  K A B = (KA)B = A$. In the next figure the combinator $\displaystyle  K$ (in green) appears as half of the 2-zipper, with one arrow and one termination gate added (in blue).

 \vspace{.5cm}

\centerline{ \includegraphics[width=80mm]{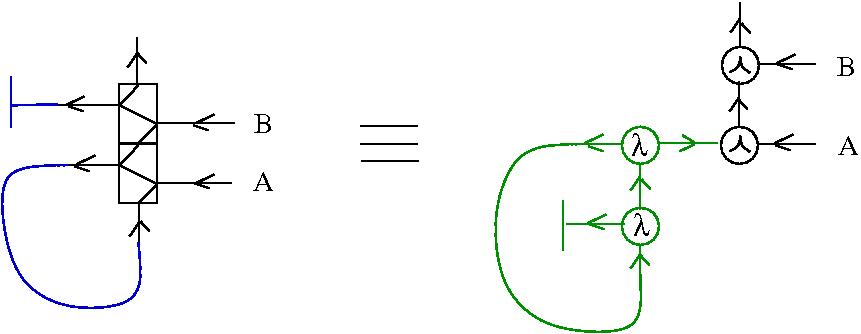}}

\vspace{.5cm}
After opening the zipper we  obtain a pair made by $\displaystyle  A$   and $\displaystyle  B$ which gets the termination gate on top of it. A global pruning move sends $B$ to the trash bin. 

Finally, the combinator $\displaystyle  S = \lambda xyz. ((xz)(yz))$ satisfies $\displaystyle  SABC = ((SA)B)C = (AC)(BC)$. The combinator $\displaystyle  S$ (in green) appears to be made by half of the 3-zipper, with some arrows and also with a "diamond" added (all in blue). Interestingly, the diamond looks alike the ones from the emergent algebra sector, definition \ref{defemers}. 

 \vspace{.5cm}

\centerline{ \includegraphics[width=80mm]{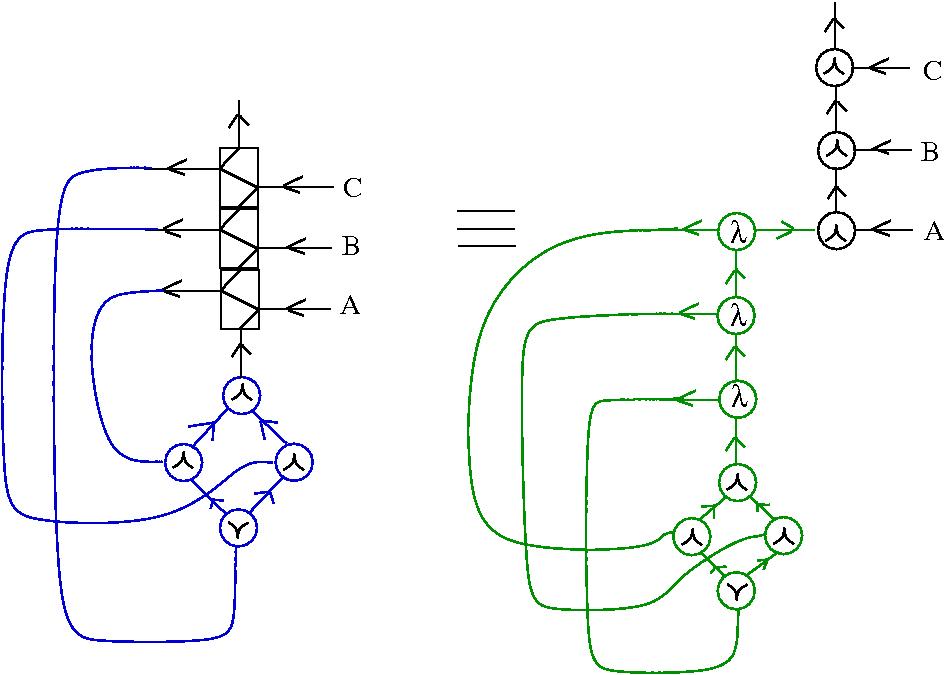}}

\vspace{.5cm}

Expressed with the help of zippers, the relation $SKK=I$ appears like this. 

 \vspace{.5cm}

\centerline{ \includegraphics[width=80mm]{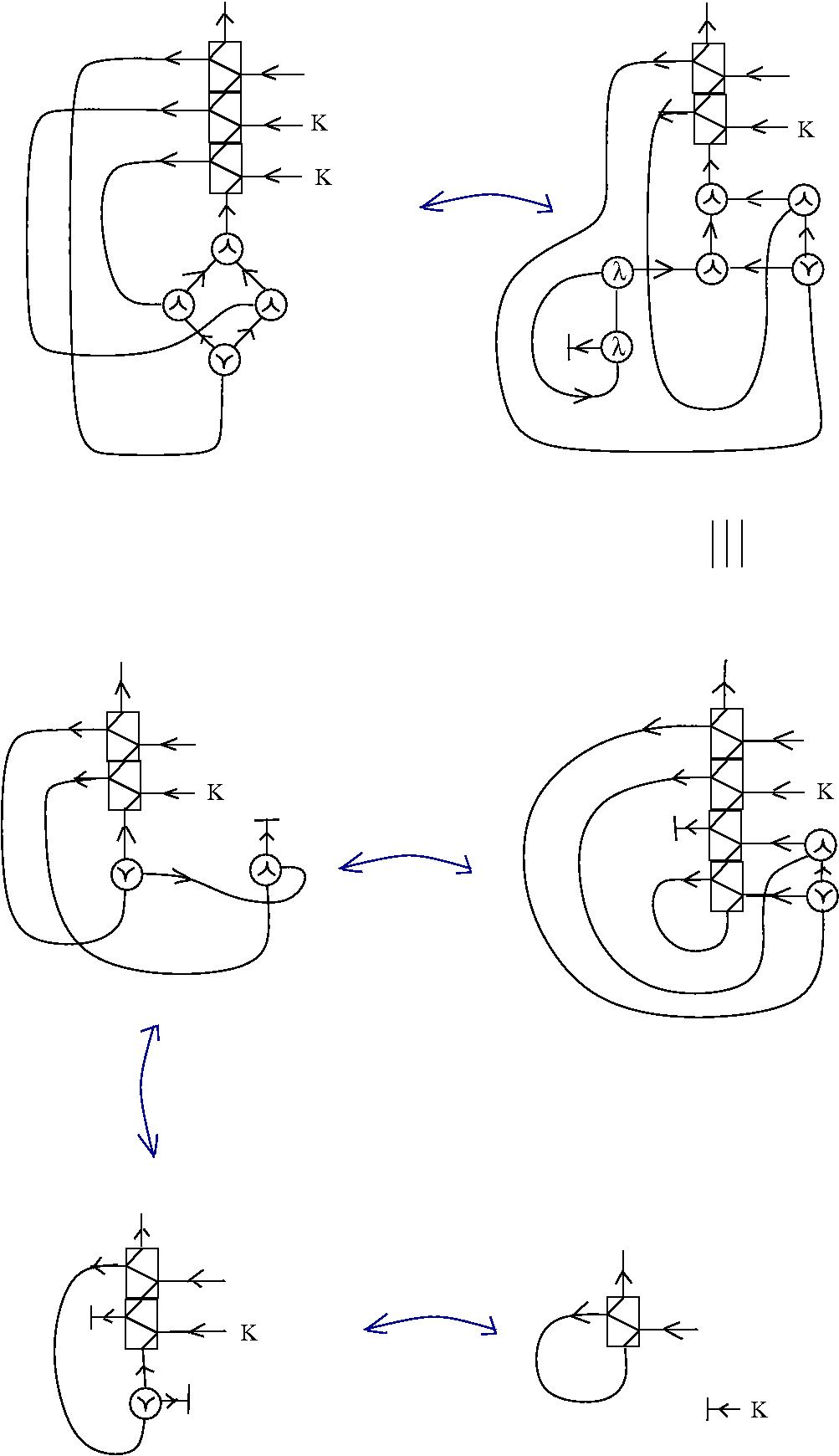}}

\vspace{.5cm}

\paragraph{Lists and currying.}
With the help of zippers, we may enhance the procedure of turning grafting into the application operation. We have a graph $\displaystyle  A \in GRAPH$ which has one output and several inputs. 

 \vspace{.5cm}

\centerline{ \includegraphics[width=20mm]{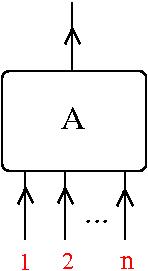}}

\vspace{.5cm}

We use an n-zipper in order to clip the inputs with the output.

 \vspace{.5cm}

\centerline{ \includegraphics[width=30mm]{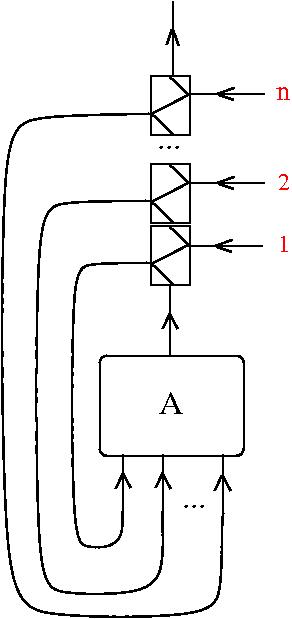}}

\vspace{.5cm}

This graph is, in fact, the following one.

 \vspace{.5cm}

\centerline{ \includegraphics[width=40mm]{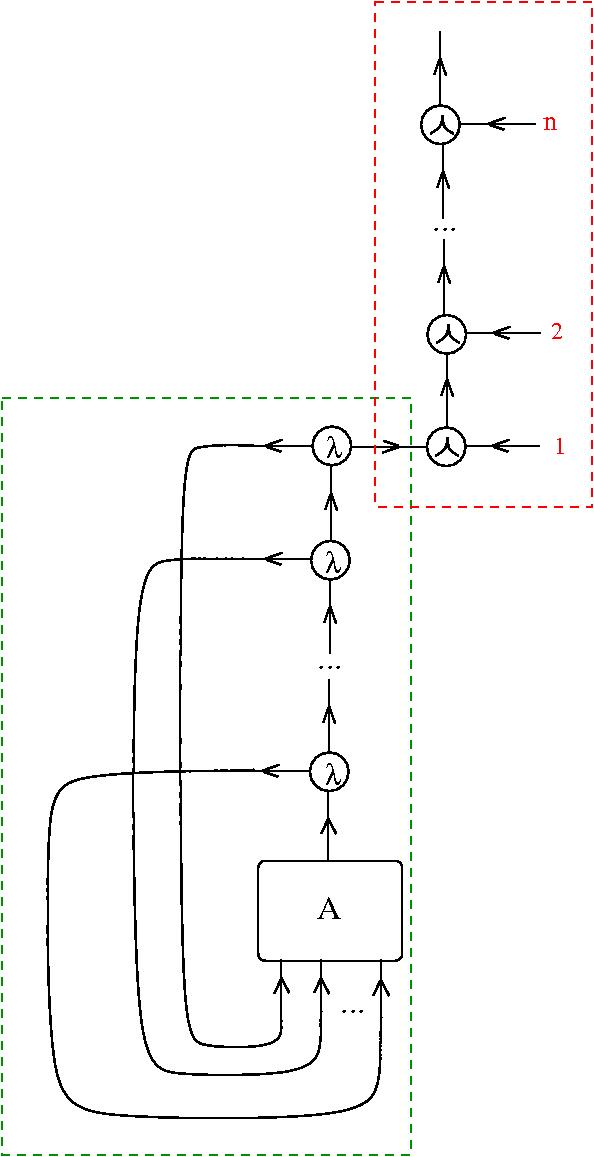}}

\vspace{.5cm}

We may interpret the graph inside the green dotted rectangle as the currying of $\displaystyle  A$, let's call him $\displaystyle  Curry(A)$.  This graph has only one output and no inputs.  The graph inside the red dotted rectangle is almost a list. We shall transform it into a list by using again a zipper and one graphic beta move.

 \vspace{.5cm}

\centerline{ \includegraphics[width=80mm]{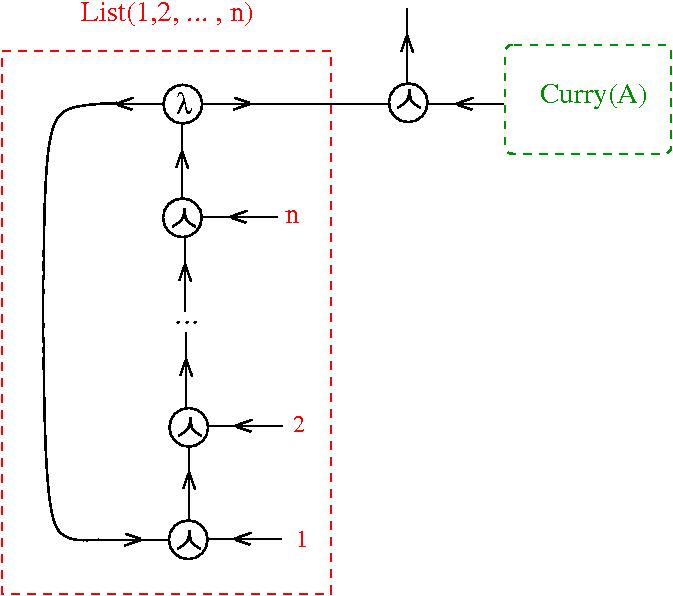}}

\vspace{.5cm}

\paragraph{Packing arrows.} We may pack several arrows into one. I describe first the case of two arrows. 
We start from the following sequence of three graphic beta moves.

 \vspace{.5cm}

\centerline{ \includegraphics[width=60mm]{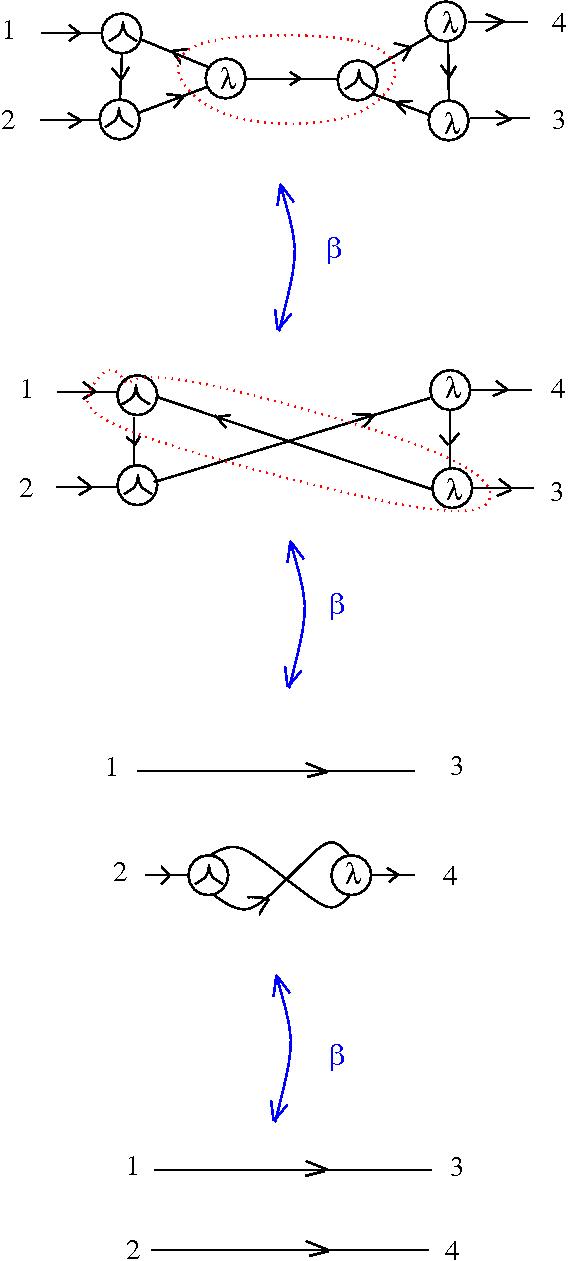}}

\vspace{.5cm}

With words, this figure means: we pack the 1, 2, entries into a list, we pass it trough one arrow then we unpack the list into the outputs 3, 4. This packing-unpacking trick may be used of course for more than a pair of arrows, in obvious ways, therefore it is not a restriction of generality to  write only about two arrows.

We may apply the trick to a  pair of graphs  $\displaystyle  A$ and $\displaystyle  B$, which are connected by a pair of arrows, like in the following figure.

\vspace{.5cm}

\centerline{ \includegraphics[width=80mm]{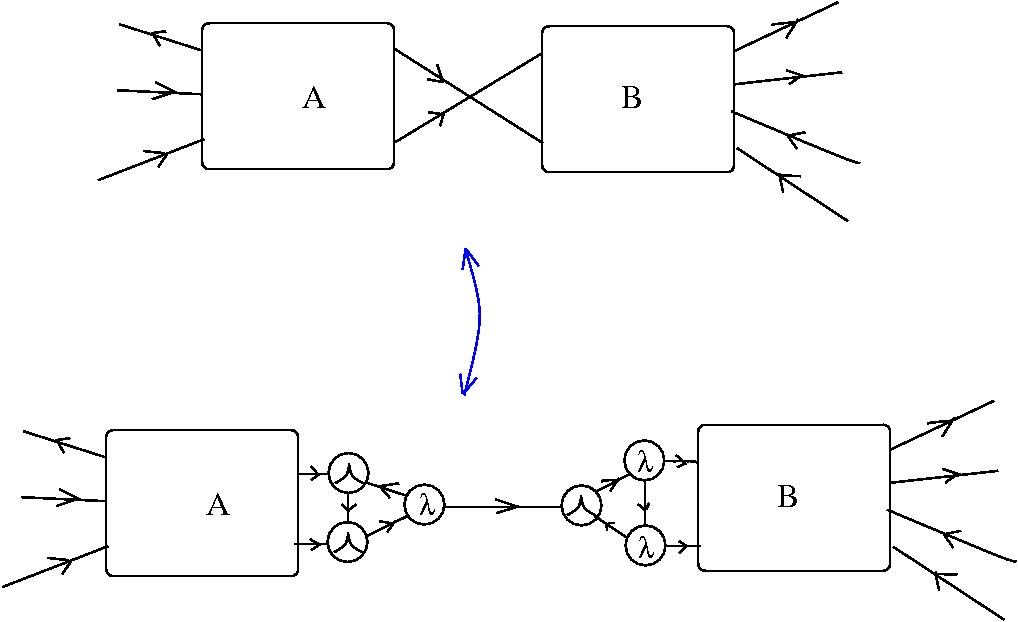}}

\vspace{.5cm}

With the added packing and unpacking triples of gates, the graphs $\displaystyle  A$, $\displaystyle  B$ are interacting only by the intermediary of one arrow.

In particular, we may use this trick for the elementary gates of abstraction and application,  transforming them into graphs with one input and one output, like this:

\vspace{.5cm}

\centerline{ \includegraphics[width=80mm]{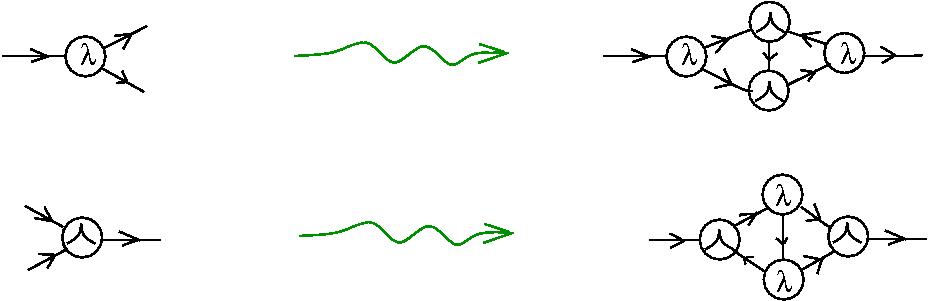}}

\vspace{.5cm}

If we use the elementary gates transformed into graphs with one input and one output, the graphic beta move becomes this almost algebraic, 1D rule:

\vspace{.5cm}

\centerline{ \includegraphics[width=80mm]{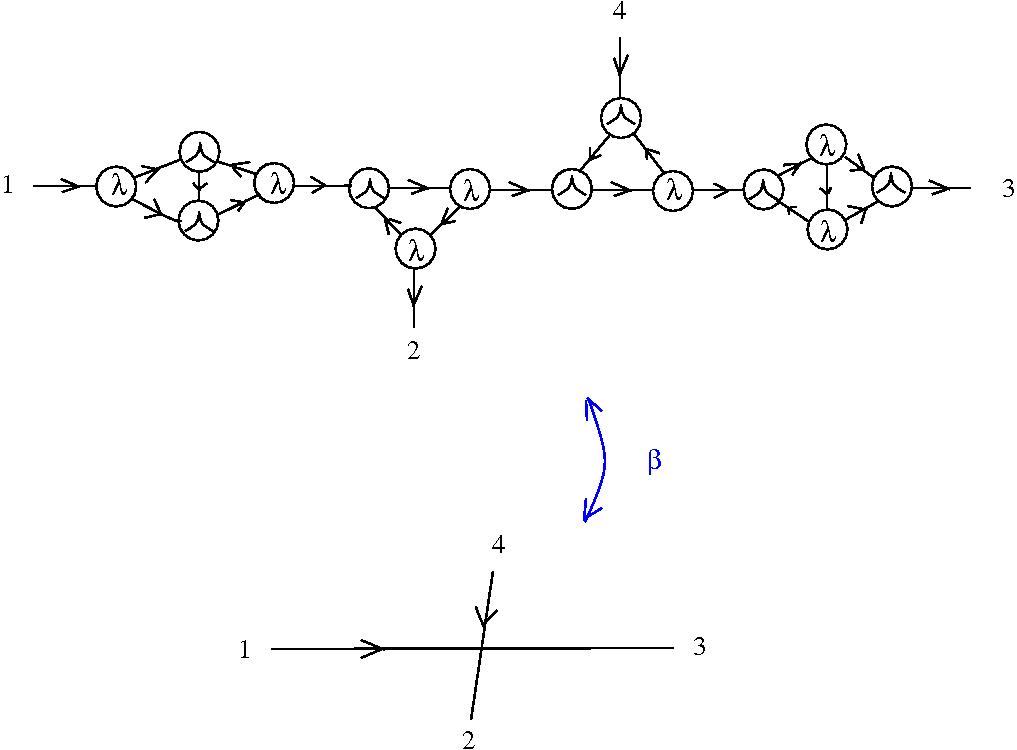}}

\vspace{.5cm} 

With such procedures, we may transform any graph in $GRAPH$ into a 1D string of graphs, consisting of transformed elementary graphs and packers and un-packers of arrows, which could be used, in principle, for transforming graphic lambda calculus into a text programming language.  

\section{Emergent algebras}
\label{semer}

Emergent algebras \cite{buligairq} \cite{buligabraided} are a distillation of differential calculus in metric spaces with dilations \cite{buligadil1}. This class of metric spaces contain the "classical" riemannian manifolds, as well as fractal like spaces as Carnot groups or, more general, sub-riemannian or Carnot-Carath\'eodory spaces, Bella\"{\i}che \cite{bell}, Gromov \cite{gromovsr}, endowed with an intrinsic differential calculus based on some variant of the Pansu derivative \cite{pansu}. 

In \cite{buligadil1} section 4 “Binary decorated trees and dilatations”, I propose a formalism for making easy various calculations with dilation structures. This formalism works with moves acting on binary decorated trees, with the leaves decorated with elements of a metric space. 

Here is an example of the formalism. The moves are (with same names as those used in graphic lambda calculus, see the explanation further):

\vspace{.5cm}

\centerline{\includegraphics[width=100mm]{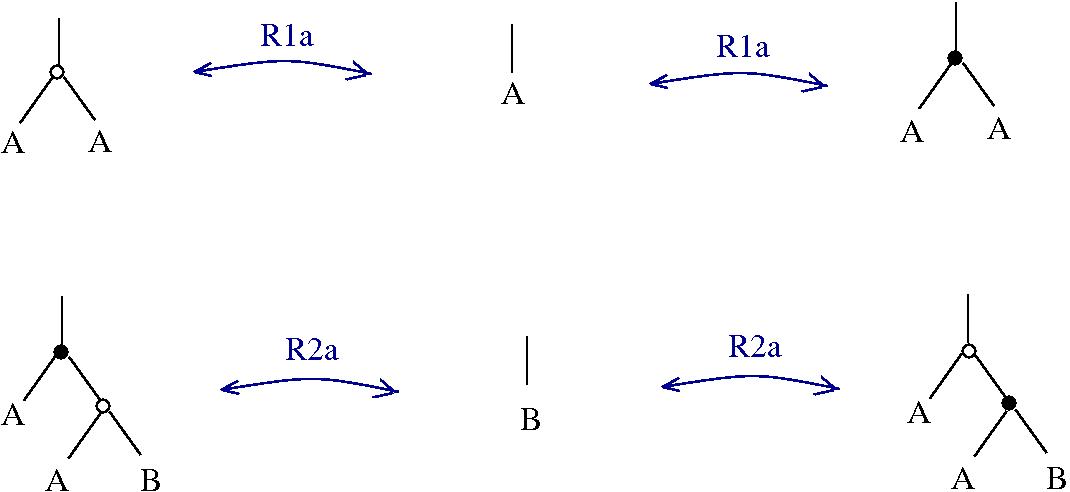}}

\vspace{.5cm}

Define the following graph (and think about it as being the graphical representation of an operation $u+v$ with respect to the basepoint $x$): 

\vspace{.5cm}

\centerline{\includegraphics[width=80mm]{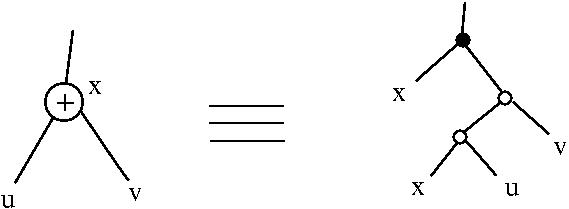}}

\vspace{.5cm}

Then, in the binary trees formalism I can prove, by using the moves R1a, R2a, the following "approximate" associativity relation (it is approximate because there appear a basepoint which is different from $x$, but which, in the geometric context of spaces with dilations, is close to $x$): 

\vspace{.5cm}

\centerline{\includegraphics[width=80mm]{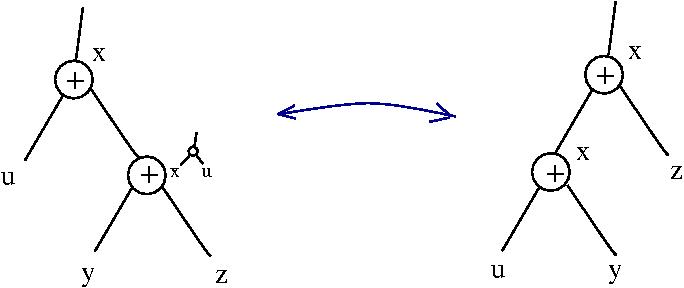}}

\vspace{.5cm}

 It was puzzling that in fact the formalism worked without needing to know which metric space is used. Moreover,  reasoning  with moves acting on binary trees gave proofs of generalizations of results from sub-riemannian geometry, while classical proofs involve elaborate calculations with pseudo-differential operators. At a close inspection it looked like somewhere in the background there is an abstract nonsense machine which is just applied to the particular case of sub-riemannian spaces.

In this paper I shall take the following pure algebraic definition of an emergent algebra (compare with definition 5.1 \cite{buligairq}), which is a stronger version of the definition 4.2 \cite{buligabraided} of a $\Gamma$ idempotent right quasigroup, in the sense that here I define a $\Gamma$ idempotent quasigroup.  

\begin{definition} Let $\Gamma$  be a commutative group with neutral element denoted by $1$ and operation denoted multiplicatively. A $\Gamma$ idempotent quasigroup is a set $X$ endowed with a family of operations $\circ_{\varepsilon}: X \times X \rightarrow X$,  indexed by $\varepsilon \in \Gamma$, such that:
\begin{enumerate}
\item[-] For any $\varepsilon \in \Gamma \setminus \left\{ 1\right\}$ the pair $\displaystyle (X, \circ_{\varepsilon})$ is an idempotent quasigroup, i.e. for any $a,b \in X$  the equations $\displaystyle x \circ_{\varepsilon} a = b$ and $a \circ_{\varepsilon} x = b$ have unique solutions and moreover $\displaystyle x \circ_{\varepsilon} x = x$ for any $x \in X$, 
\item[-] The operation $\circ_{1}$ is trivial: for any $x,y \in X$ we have $x \circ_{1} y = y$, 
\item[-] For any $x, y \in X$ and any $\varepsilon, \mu \in \Gamma$ we have:  $x \circ_{\varepsilon} ( x \circ_{\mu} y) = x \circ_{\varepsilon \mu} y$.
\end{enumerate}
\label{defirq}
\end{definition}

Here are  some examples of $\Gamma$ idempotent quasigroups.

\paragraph{Example 1.}  Real (or complex) vector spaces: let $X$ be a real (complex)  vector space, $\Gamma = (0,+\infty)$ (or $\displaystyle \Gamma = \mathbb{C}^{*}$), with multiplication as operation. We define, for any $\varepsilon \in \Gamma$  the following quasigroup operation: $\displaystyle x \circ_{\varepsilon} y = (1-\varepsilon) x + \varepsilon y$.  These operations give to $X$ the structure of a $\Gamma$ idempotent quasigroup.  Notice that $\displaystyle x \circ_{\varepsilon}y $ is the dilation based at $x$, of coefficient $\varepsilon$, applied to $y$.

\paragraph{Example 2.}     Contractible groups: let $G$ be a group endowed with a group morphism $\phi: G \rightarrow G$. Let $\Gamma = \mathbb{Z}$ with the operation of addition of integers (thus we may adapt definition \ref{defirq} to this example by using "$\varepsilon + \mu$" instead of "$\varepsilon \mu$" and "$0$" instead of "$1$" as the name of the neutral element of $\Gamma$).  For any $\varepsilon \in \mathbb{Z}$ let $\displaystyle x \circ_{\varepsilon} y = x \phi^{\varepsilon}(x^{-1} y)$.  This a $\mathbb{Z}$ idempotent quasigroup. The most interesting case  is the one when $\phi$ is an uniformly contractive automorphism of a  topological group $G$. The structure of these groups is an active exploration area, see for example \cite{glockner}  and the bibliography therein. A fundamental result here is Siebert \cite{siebert}, which gives a characterization of topological connected contractive locally compact groups as being nilpotent Lie groups endowed with a one parameter family of dilations, i.e. almost Carnot groups.

\paragraph{Example 3.}  A group with an invertible self-mapping $\phi: G \rightarrow G$  such that $\phi(e) =e$, where $e$ is the identity of the group $G$. It looks like  Example 2 but it shows that  there is no need for $\phi$ to be a group morphism.

\paragraph{Local versions.}   We may accept that there is a way (definitely needing care to well formulate, but intuitively clear) to define a local version of the notion of a $\Gamma$  idempotent quasigroup. With such a definition, for example, a convex subset of a real vector space gives a local $(0,+\infty)$ idempotent quasigroup (as in Example 1) and a neighbourhood of the identity of a topological group $G$, with an identity preserving, locally defined invertible self map (as in Example 3) gives a $\mathbb{Z}$ local idempotent quasigroup.

\paragraph{Example 4.}   A particular case of Example 3  is a Lie group $G$ with the operations  defined for any $\varepsilon \in (0,+\infty)$ by $x \circ_{\varepsilon} y = x \exp ( \varepsilon \log (x^{-1} y) )$. 

\paragraph{Example 5.}  A less symmetric example is the one of $X$ being a riemannian manifold, with associated operations  defined for any $\varepsilon \in (0,+\infty)$ by $x \circ_{\varepsilon}y = \exp_{x}( \varepsilon \log_{x}(y))$, where $\exp$ is the metric exponential. 

\paragraph{Example 6.}  More generally, any metric space with dilations is a local idempotent (right) quasigroup.

\paragraph{Example 7.}   One parameter deformations of quandles. A quandle is a self-distributive quasigroup. Take now a one-parameter family of quandles (indexed by $\varepsilon \in \Gamma$) which satisfies moreover points 2. and 3. from definition \ref{defirq}. What is interesting about this example is that quandles appear as decorations of knot diagrams \cite{fennrourke} \cite{joyce}, which are preserved by the Reidemeister moves (more on this in the section \ref{secbraid}).  At closer examination, examples 1, 2 are  particular cases of one parameter quandle deformations!

I  define now the operations of approximate sum and approximate difference associated to a $\Gamma$  idempotent quasigroup.

\begin{definition}
For any $ \varepsilon \in \Gamma$ we give the following names to several combinations of operations of emergent algebras:
\begin{enumerate}
\item[-] the approximate sum operation is $\displaystyle  \Sigma^{x}_{\varepsilon} (u,v) = $  $\displaystyle x \bullet_{\varepsilon} ((x \circ_{\varepsilon} u) \circ_{\varepsilon} v)$,
\item[-] the approximate difference operation is  $\displaystyle \Delta^{x}_{\varepsilon} (u,v) = (x \circ_{\varepsilon} u) \bullet_{\varepsilon} (x \circ_{\varepsilon} v)$, 
\item[-] the approximate inverse operation is  $\displaystyle inv^{x}_{\varepsilon} u = (x \circ_{\varepsilon} u) \bullet_{\varepsilon} x$.
\end{enumerate}
\end{definition}

Let's see what the approximate sum operation is, for example 1. 
$$\Sigma^{x}_{\varepsilon}(u,v) = u(1-\varepsilon) -x + v $$
It is clear that, as $\varepsilon$ converges to $0$, this  becomes the operation of addition in the vector space with $x$ as neutral element, so it might be said that is the operation of addition of vectors in the tangent space at $x$, where $x$ is seen as an element of the affine space constructed over the vector space from example 1. 

This is a general phenomenon, which becomes really interesting in non-commutative situations, i.e. when applied to examples from the end of the provided list. 

These approximate operations have many algebraic properties which can be found by the abstract nonsense of manipulating binary trees.

Another construction which can be done in emergent algebras is the one of taking finite differences (at a high level of generality, not bonded to vector spaces). 

\begin{definition}
Let $A: X \rightarrow X$ be a function (from $X$ to itself, for simplicity). The finite difference function associated to $A$, with respect to the emergent algebra over $X$, at a point $x \in X$ is the following.  
$$T_{\varepsilon}^{x} A : X \rightarrow X \quad , \quad T_{\varepsilon}^{x} A (u)  = A(x) \bullet_{\varepsilon} \left( A \left( x \circ_{\varepsilon} u \right) \right)$$
\label{defpan}
\end{definition}

For example 1, the finite difference has the expression: 
$$T^{x}_{\varepsilon} A( u - x)  =  A(x) + \frac{1}{\varepsilon} \left( A(x + \varepsilon u) - A(x) \right)$$
which is a finite difference indeed. In more generality, for example 2 this definition leads to the Pansu derivative \cite{pansu}. 

Finite differences as defined here behave like discrete versions of derivatives. Again, the proofs consist in manipulating well chosen binary trees. 

All this can be formalized in graphic lambda calculus, thus transforming the proofs into computations inside graphic lambda calculus. 

I shall not insist more on this, with the exception of describing  the emergent algebra sector of graphic lambda calculus. 

\begin{definition} For any $\displaystyle  \varepsilon \in \Gamma$ , the following graphs in $\displaystyle  GRAPH$ are introduced:
\begin{enumerate}
\item[-] the approximate sum graph $\displaystyle  \Sigma_{\varepsilon}$ 
\end{enumerate}

\vspace{.5cm}

\centerline{\includegraphics[width=30mm]{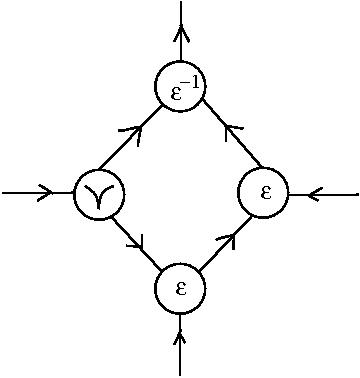}}

\vspace{.5cm}

 \begin{enumerate}
	 \item[-]the approximate difference graph $\displaystyle  \Delta_{\varepsilon}$ 
\end{enumerate}
\vspace{.5cm}

\centerline{\includegraphics[width=25mm]{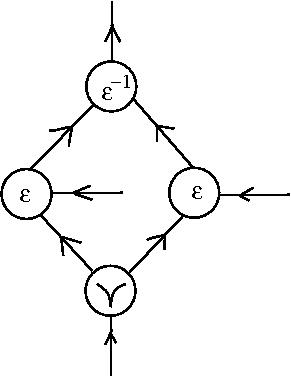}}

\vspace{.5cm}

 \begin{enumerate}
	 \item[-]the approximate inverse graph $\displaystyle  inv_{\varepsilon}$ 
\end{enumerate}
\vspace{.5cm}

\centerline{\includegraphics[width=15mm]{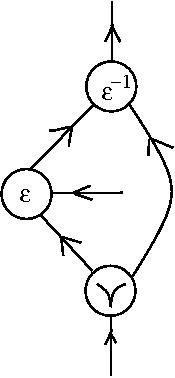}}

\vspace{.5cm}
\label{defemers}
\end{definition}

Let $\displaystyle  A$ be a set of symbols $\displaystyle  a, b, c, ...$.   (These symbols will play the role of  scale parameters going to $\displaystyle  0$.) With $\displaystyle  A$ and with the abelian group $\displaystyle  \Gamma$ we construct a larger abelian group, call it $\displaystyle  \bar{\Gamma}$, which is generated by $\displaystyle  A$ and by $\displaystyle  \Gamma$.

Now we introduce the emergent algebra sector (over the set $\displaystyle  A$).

\begin{definition} $\displaystyle  EMER(A)$  is the subset of $\displaystyle  GRAPH$ (over the group $\displaystyle  \bar{\Gamma}$)  which is generated by the following list of gates:
 \begin{enumerate}
	 \item[-]arrows and loops, 
	 \item[-]$\displaystyle  \Upsilon$ gate and the gates $\displaystyle  \bar{\varepsilon}$ for any $\displaystyle  \varepsilon \in \Gamma$, 
	 \item[-]the approximate sum gate $\displaystyle  \Sigma_{a}$ and the approximate difference gate $\displaystyle  \Delta_{a}$, for any $\displaystyle  a \in A$, 
\end{enumerate}
with the operations of linking output to input arrows  and with the following list of moves:
 \begin{enumerate}
	 \item[-]FAN-OUT moves
	 \item[-]emergent algebra moves  for the group $\displaystyle  \bar{\Gamma}$, 
	 \item[-]<pruning moves. 
\end{enumerate}
The set $\displaystyle  EMER(A)$ with the given list of moves is called the emergent algebra sector over the set $\displaystyle  A$.
\label{defemerse}
\end{definition}

The approximate inverse is not included into the list of generating gates.  That is because we can prove easily that for any $\displaystyle  a \in A$ we have $\displaystyle  inv_{a} \in EMER(A)$.  (If $\displaystyle  \varepsilon \in \Gamma$ then  we trivially have $\displaystyle  inv_{\varepsilon} \in EMER(A)$ because it is constructed from emergent algebra gates decorated by elements in $\displaystyle  \Gamma$, which are on the list of generating gates.) Here is the proof: we start with the approximate difference $\displaystyle  \Delta_{a}$ and with an $\displaystyle  \Upsilon$ gate and we arrive to the approximate inverse $\displaystyle  inv_{a}$ by a sequence of moves, as follows.

\vspace{.5cm}

\centerline{\includegraphics[width=80mm]{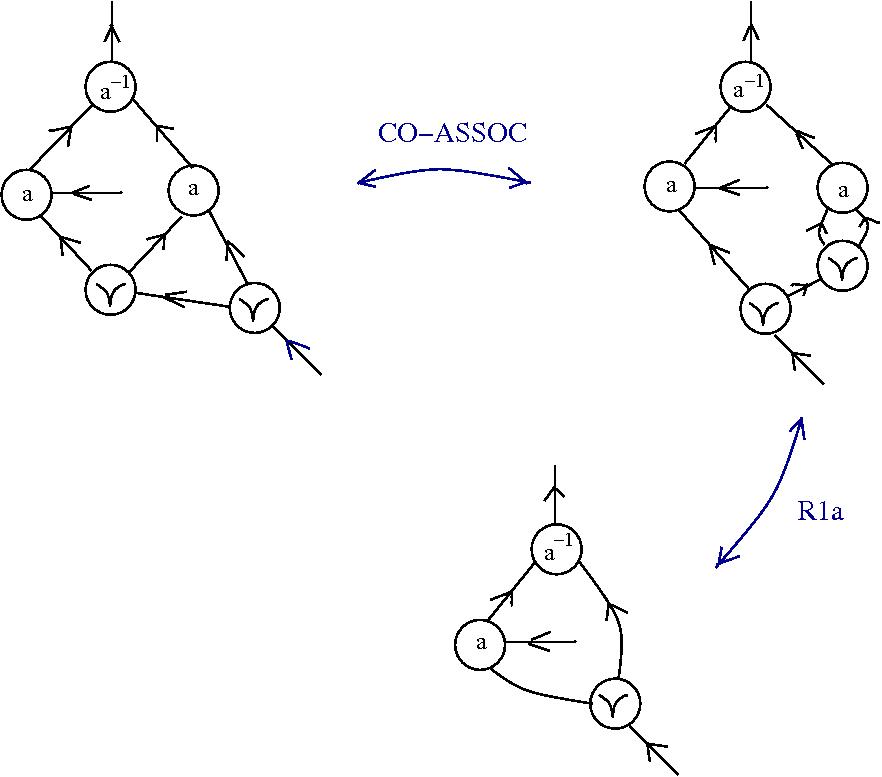}}

\vspace{.5cm}

We proved the following relation for emergent algebras: $\displaystyle  \Delta^{x}_{a} (u,x) = \, inv^{x}_{a} u$.  This relation appears as a computation in graphic lambda calculus. 

As for the finite differences, we may proceed as this. 

\begin{definition}
A graph $A \in GRAPH$, with one input and one output distinguished, is computable with respect to the group $\displaystyle \bar{Gamma}$ if the following graph 

\vspace{.5cm}

\centerline{\includegraphics[width=25mm]{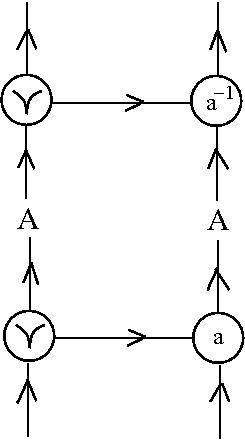}}

\vspace{.5cm}
can be transformed by the moves from graphic lambda calculus into a graph which is made by assembling: 
\begin{enumerate}
\item[-] graphs from $EMER(A)$,
\item[-] gates $\lambda$, $\curlywedge$ and $\top$. 
\end{enumerate}
\label{defemercomp}
\end{definition}

It would be interesting to mix the emergent algebra sector with the lambda calculus sector (in a sense this is already suggested in definition \ref{defemercomp}). At first view, it seems that the emergent algebra gates $\displaystyle \bar{\varepsilon}$ are operations which are added to the lambda calculus operations, the latter being more basic than the former. I think this is not the case. In \cite{lambdascale} theorem 3.4, in the formalism of lambda-scale calculus (graphic lambda calculus is a visual variant of this), I show on the contrary that the emergent algebra gates could be applied to lambda terms and the result is a collection, or hierarchy of lambda calculi, organized into an emergent algebra structure. This is surprising, at least for the author, because the initial goal of introducing lambda-scale calculus was to mimic lambda calculus with emergent algebra operations.

\section{Crossings}
\label{secbraid}

In this section we discuss about tangle diagrams and graphic lambda calculus. 

An oriented tangle is a collection of wired in 3D space, more precisely it is an embedding of a oriented one dimensional manifold in 3D space. Two tangles are the same up to topological deformation of the 3D space.  An oriented tangle diagram  is, visually, a projection of a tangle, in general position,  
on a plane. More specifically, an oriented tangle diagram is a globally planar oriented graph with 4-valent nodes which represent crossings of wires (as seen in the projection), along with supplementary information about which wire passes over the respective crossing. A locally planar tangle diagram is an oriented graph which satisfies the previous description, with the exception that it is only locally planar. Visually, a locally planar tangle diagram looks like an ordinary one, excepting that there may be crossings of edges of the graph which are not tangle crossings (i.e. nodes of the graph).

The purpose of this section is to show that we can "simulate" tangle diagrams with graphic lambda calculus. This can be expressed more precisely in two ways. The first way  is that we can define "crossing macros" in graphic lambda calculus, which are certain graphs which play the role of crossings in a tangle diagram (i.e. we can express the Reidemeister moves, described further, as compositions of moves from graphic lambda calculus between such graphs). The second way is to say that to any tangle diagram we can associate a graph in $GRAPH$ such that to any Reidemeister move is associated a certain composition of moves from graphic lambda calculus. 

Meredith ad Snyder \cite{mersny} achieve this goal with the pi-calculus instead of graphic lambda calculus. Kauffman, in the second part of \cite{kauf}, associates tangle diagrams to combinators and writes about "knotlogic".

\paragraph{Oriented Reidemeister moves.} 
Two tangles are the same, up to topological equivalence, if and only if any tangle diagram of one tangle can be transformed by a finite sequence of Reidemeister moves into a tangle diagram of the second tangle. 
The oriented Reidemeister moves are the following (I shall use the same names as Polyak \cite{polyak}, but with the letter $\displaystyle \Omega$  replaced by the letter $\displaystyle R$):
\begin{enumerate}
\item[-] four oriented Reidemeister moves of type 1:
\vspace{.5cm}

\centerline{\includegraphics[width=100mm]{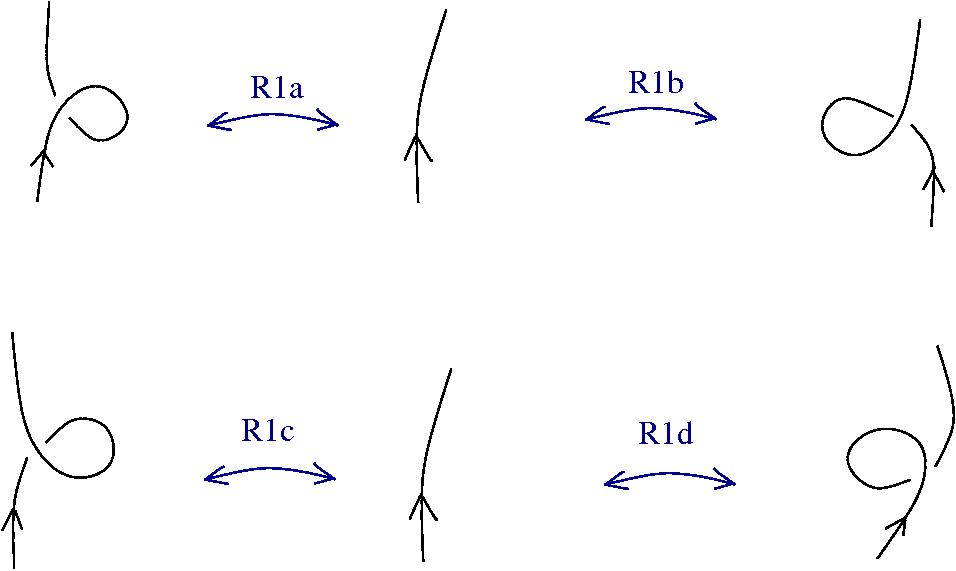}}

\vspace{.5cm}

\item[-] four oriented Reidemeister moves of type 2:

\vspace{.5cm}

\centerline{\includegraphics[width=100mm]{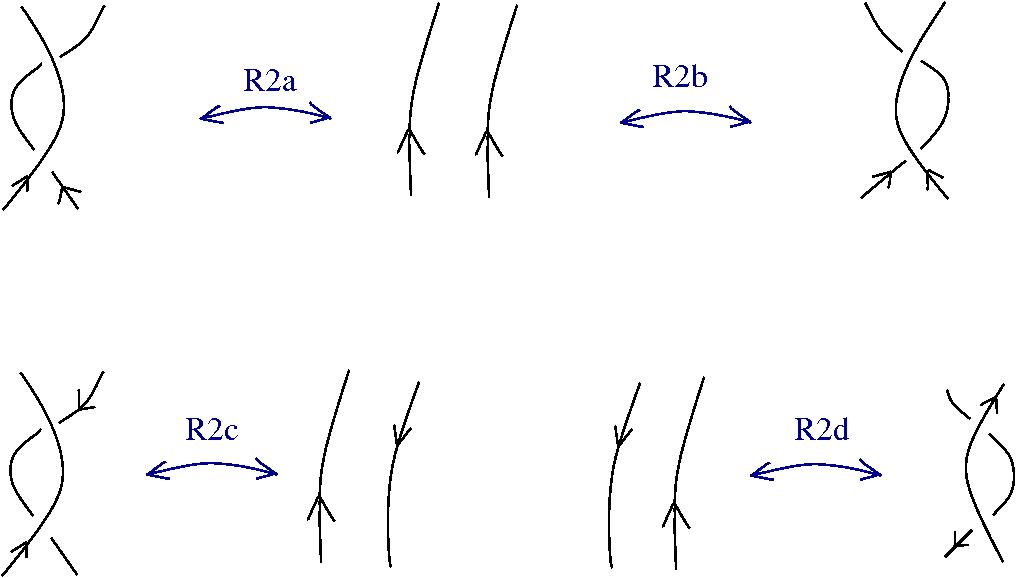}}

\vspace{.5cm}

\item[-] eight oriented Reidemeister moves of type 3:

\vspace{.5cm}

\centerline{\includegraphics[width=100mm]{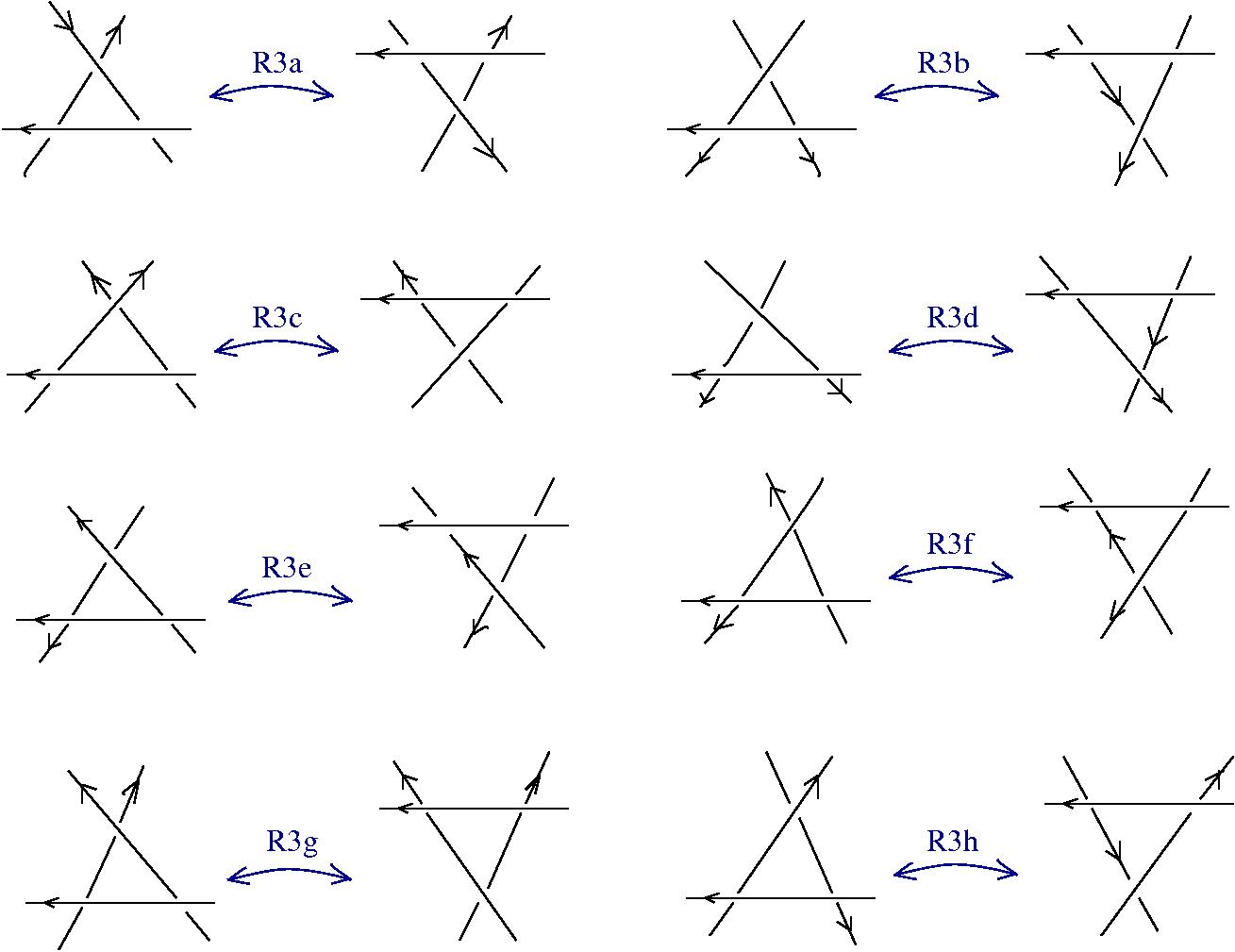}}

\vspace{.5cm}
\end{enumerate}

\paragraph{Crossings from emergent algebras.} In section \ref{semer}, example 7, it is mentioned that there is a connection between tangle diagrams and emergent algebras, via the notion of a quandle. Quandles are self-distributive idempotent quasigroups, which were invented as decorations of the arrows of a tangle diagram, which are invariant with respect to the Reidemeister moves.

Let us define the emergent algebra crossing macros. (We can choose to neglect the $\varepsilon$ decorations of the crossings, or, on the contrary, we can choose to do like in definition \ref{defemerse} of the emergent algebra sector, namely to add a set $A$ to the group $\Gamma$ and use even more nuanced decorations for the crossings.)

\vspace{.5cm}

\centerline{\includegraphics[width=90mm]{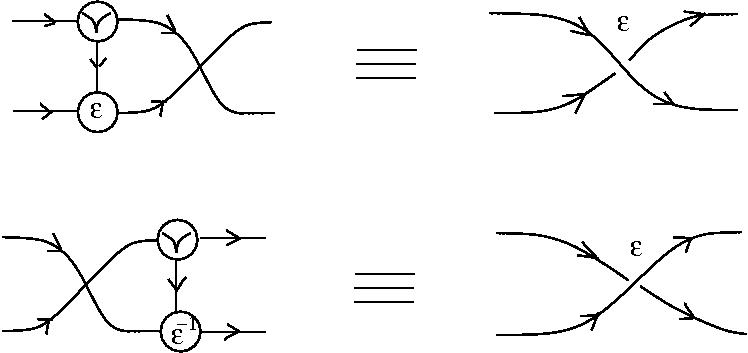}}

\vspace{.5cm}

In \cite{buligachora}, sections 3-6 are dedicated to the use of these crossings for exploring  emergent algebras and spaces with dilations. All constructions and reasonings from there can be put into the graphic lambda calculus formalism. Here I shall explain only some introductory facts. 

Let us associate to any locally planar tangle diagram $T$ a graph in $[T] \in GRAPH$, called the translation of $T$, which is obtained by replacing the crossings with the emergent crossing macros (for a fixed $\varepsilon$). Also,  to any Reidemeister move we associate it's translation in graphic lambda calculus, consisting in a local move between the translations of the LHS and RHS tangles which appear in the respective move. (Note: these translations are not added to the moves which define graphic lambda calculus.)

\begin{theorem}
The translations of all oriented Reidemeister moves of type 1 and 2 can be realized as  sequences of the following moves from graphic lambda calculus: emergent algebra moves (R1a, R1b, R2, ext2),  fan-out moves (i.e. CO-COMM, CO-ASSOC, global FAN-OUT) and pruning moves. More precisely the translations of the Reidemeister moves R1a, R1b are, respectively, the graphic lambda calculus moves R1a, R1b, modulo fan-out moves. Moreover, all translations of Reidemeister moves of type 2 can be expressed in graphic lambda calculus with the move R2, fan-out and pruning moves. 
\label{temercross}
\end{theorem}

The proof is left to the interested reader, see however section 3.4 \cite{buligachora}. 

The fact that the Reidemeister moves of type 3 are not true for (the algebraic version of) the emergent algebras, i.e. that the translations of those cannot be expressed as a sequence of moves from graphic lambda calculus, is a feature of the formalism and not a weakness. This is explained in detail in sections 5, 6 \cite{buligachora}, but unfortunately at the moment of the writing that article the graphic lambda calculus was not available. It is an interesting goal the one of expressing the constructions from the mentioned sections as statements about the computability in the sense of definition \ref{defemercomp} of the translations of certain tangle diagrams. 

As a justification for this point of view, let us remark that all tangle diagrams which appear in the Reidemeister moves of type 3 have translations which are related to the approximate difference or approximate sum graphs from definition \ref{defemers}. For example, let's take the translation of the graph from the RHS of the move R3d and call it $D$. This graph has three inputs and three outputs. Let's then consider a graph formed by grafting three graphs $A$, $B$, $C$ at the inputs of $D$, such that $A$, $B$, $C$ are not otherwise connected. Then we can perform the following sequence of moves.

\vspace{.5cm}

\centerline{\includegraphics[width=110mm]{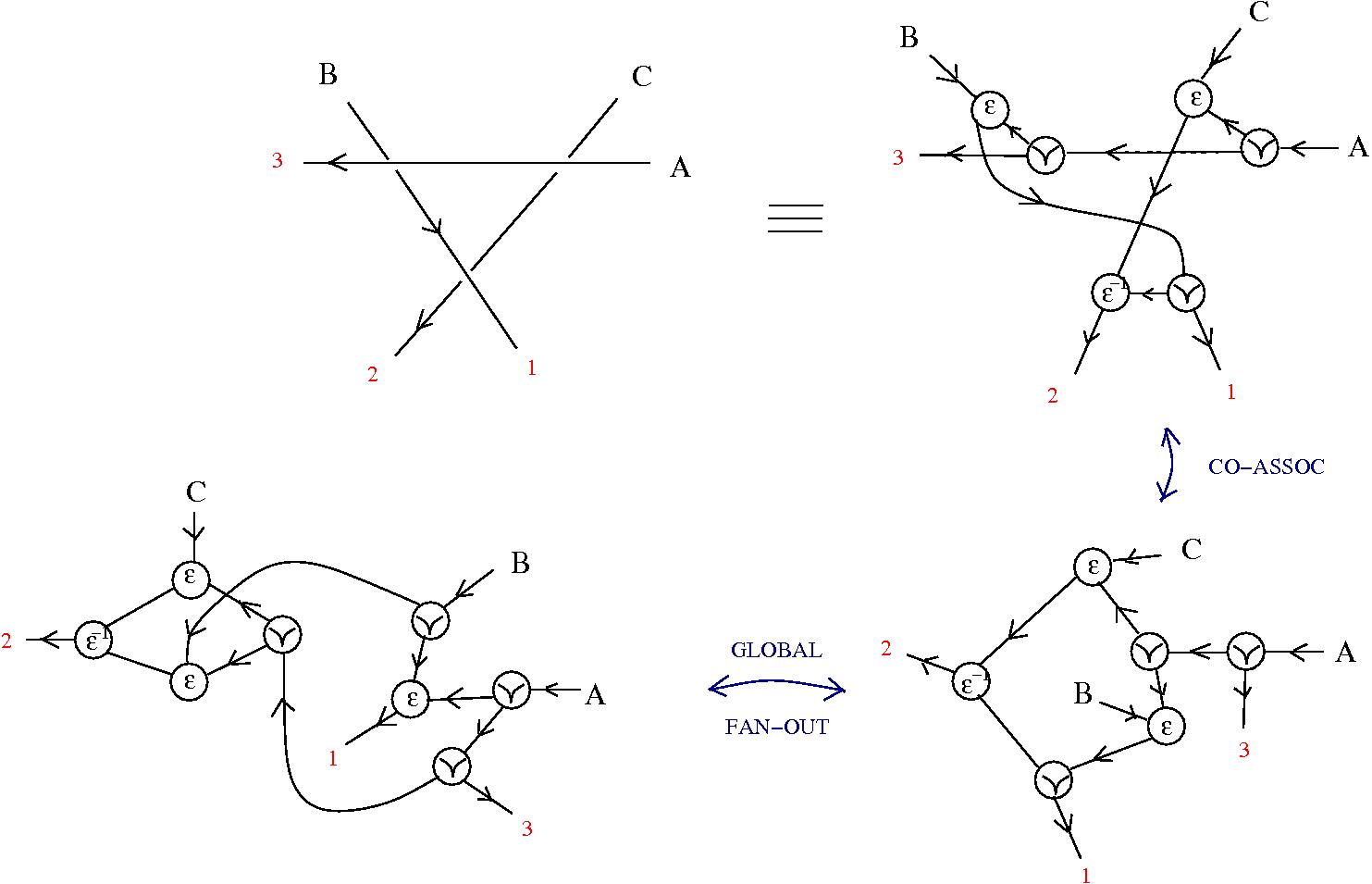}}

\vspace{.5cm}

The graph from the left lower side is formed by an approximate difference, a $\bar{\varepsilon}$ gate and several $\Upsilon$ gates. Therefore, if $A$, $B$, $C$ are computable in the sense of definition \ref{defemers} then the initial graph (the translation of the LHS of R3d with $A$, $B$, $C$ grafted at the inputs) is computable too. 

\paragraph{Graphic beta move as braiding.} 
Let us now construct crossings, in the sense previously explained, from gates coming from lambda calculus. 

\vspace{.5cm}

\centerline{\includegraphics[width=90mm]{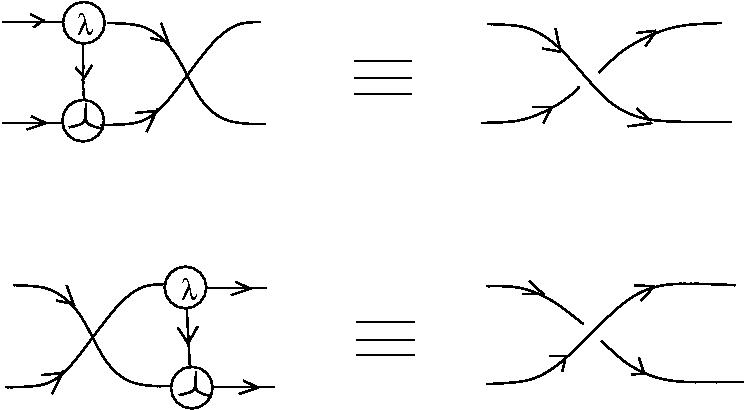}}

\vspace{.5cm}

As previously, we define translations of (locally planar) tangle diagrams into graphs in $GRAPH$. The class of locally planar tangle diagrams is out in a one-to one correspondence with a class of graphs in $GRAPH$, let us call this class $\lambda-TANGLE$. 

We could proceed in the inverse direction, namely consider the class of graphs $\lambda-TANGLE$, along with the moves: graphic beta move and elimination of loops. Then we make the (inverse) translation of graphs in $\lambda-TANGLE$ into locally planar tangle diagrams and the (inverse) translation of the graphic beta move and the elimination of loops. The following proposition explains what we obtain. 

\begin{proposition}
The class of graphs $\lambda-TANGLE$ is closed with respect to the application of the graphic beta move and of the elimination of loops. The translations of the graphic beta and elimination of loops moves are the following SPLICE 1, 2 (translation of the graphic beta move) and LOOP 1, 2 (translation of the elimination of loops) moves. 
\label{ptrans}
\end{proposition}

 \vspace{.5cm}

\centerline{\includegraphics[width=100mm]{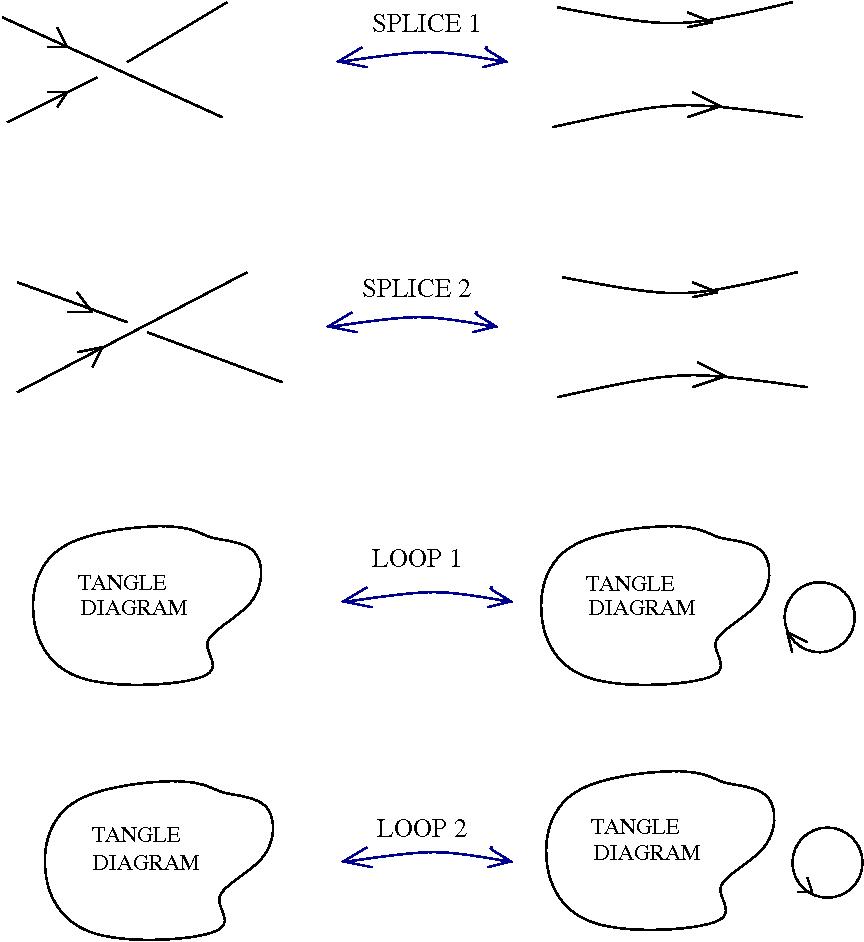}}

\vspace{.5cm}

\paragraph{Proof.} The  proposition becomes obvious if we find the translation of the graphic beta move. There is one translation for each crossing. (Likewise, there are two translations for elimination of loops, depending on the orientation of the loop which is added/erased.) \quad $\square$

\vspace{.5cm}

\centerline{\includegraphics[width=90mm]{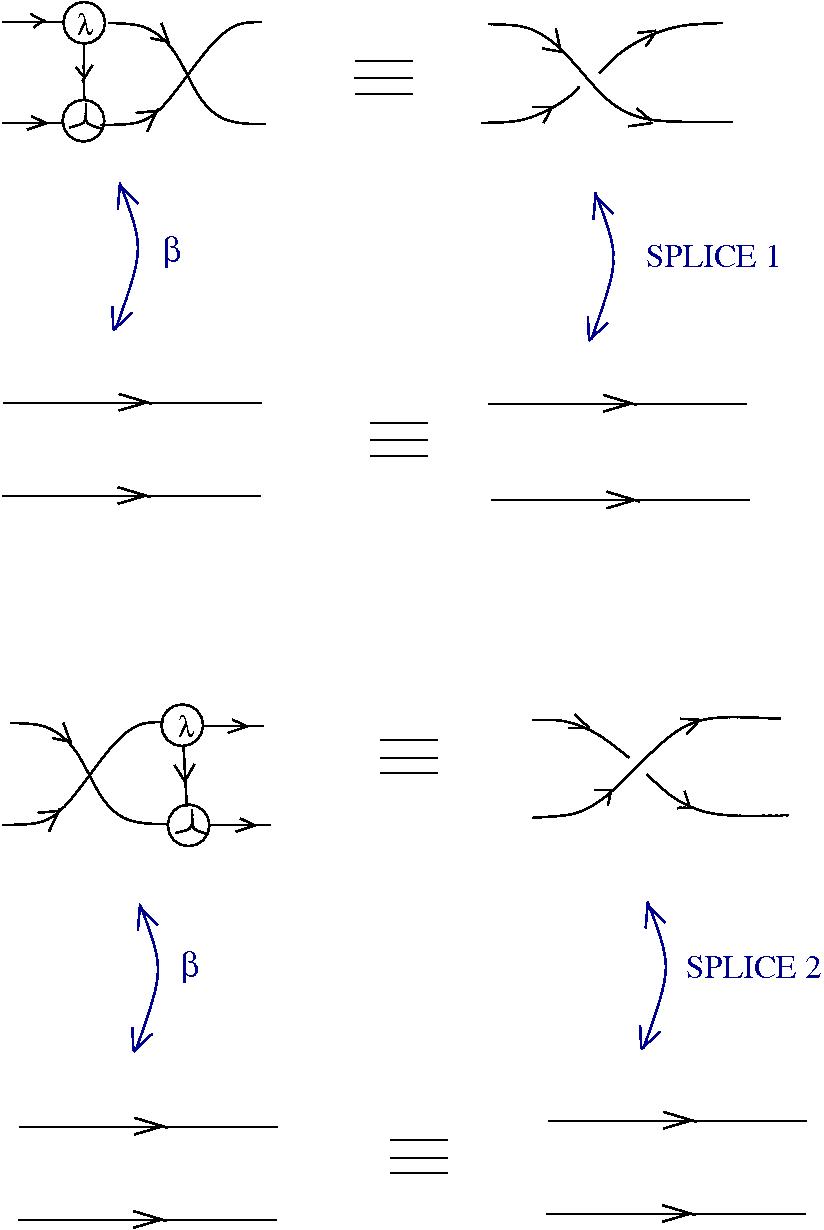}}

\vspace{.5cm}

The following theorem clarifies which are the oriented Reidemeister moves which can be expressed as sequences of graphic lambda calculus moves applied to graphs in $\lambda-TANGLE$.  
Among these moves, some are more powerful than others, as witnessed by the following

\begin{theorem}
 All the translations of the oriented Reidemeister move into moves between graphs in $\lambda-TANGLE$, excepting R2c, R2d, R3a, R3h, can be realized as sequences of graphic beta moves and elimination of loops. Moreover, the translations of moves R2c, R2d, R3a, R3h are equivalent up to graphic beta moves and elimination of loops (i.e. any of these moves, together with the graphic beta move and elimination of loops, generates the other moves from this list).
\label{threidgen}
\end{theorem}

\paragraph{Proof.} It is easy, but tedious, to verify that all the mentioned moves can be realized as sequences of SPLICE and LOOP moves. It is as well easy to verify that the moves R2c, R2d, R3a, R3h are equivalent up to SPLICE and LOOP moves. It is not obvious that the moves R2c, R2d, R3a, R3h can't be realized as a sequence of SPLICE and LOOP moves. In order to do this, we prove that  R2d can't be generated by SPLICE and LOOP. Thanks are due to Peter Kravchuk for the idea of the proof,   given in an answer to a question I asked on mathoverflow \cite{buligamo}, where I described the moves SPLICE and LOOP.

To any locally planar tangle diagram A associate it's reduced diagram R(A), which is obtained by the following procedure: first use SPLICE 1,2 from left to right for all crossings, then use LOOP 1,2 from right to left in order to eliminate all loops which are present at this stage. Notice that: 
\begin{enumerate}
\item[-] the order of application of the SPLICE  moves does not matter, because they are applied only once per crossing. There is a finite number of splices, equal to the number of crossings. Define the bag of splices SPLICE(A) to be the set of SPLICE  moves applied.
\item[-]  The same is true for the order of eliminations of loops by LOOP 1, 2. There is a finite number of loop eliminations, because the number of loops (at this stage) cannot be bigger than the number of edges of the initial diagram. Define the bag of loops LOOP(A) to be the set of all loops which are present after all splices are done.
\end{enumerate}

Let us now check that the reduced diagram does not change if one of the 4 moves is applied to the initial diagram.

Apply a SPLICE 1,2 move to the initial diagram A, from left to right, and get B. Then SPLICE(B) is what is left in the bag SPLICE(A) after taking out the respective splice. Also LOOP(B) = LOOP(A) because of the definition of bags of loops. Therefore R(A) = R(B).

Apply a SPLICE 1, 2 from right to left to A and get B. Then R(A) = R(B) by the same proof, with A, B switching places.

Apply a LOOP1, 2 from left to right to A and get B. The new loop introduced in the diagram does not participate to any crossing (therefore SPLICE(A) = SPLICE(B)), so we find it in the bag of loops of B, which is made by all the elements of LOOP(A) and this new loop. Therefore R(A) = R(B). Same goes for LOOP1, 2 applied from right to left.

Finally, remark that the reduced diagram of the LHS of the move R2d is different than the reduced diagram of the RHS of the move R2d, therefore the move R2d cannot be achieved with a sequence of splices and loops addition/elimination. \quad $\square$

\end{document}